\documentclass[prd,preprint,superscriptaddress,preprintnumbers,eqsecnum,showpacs,nofootinbib,nobibnotes]{revtex4}
\usepackage{amsfonts,amsmath,amssymb,bm,natbib}
\usepackage{graphicx}  

\newcommand{\be}{\begin{equation}}
\newcommand{\bea}{\begin{eqnarray}}
\newcommand{\ee}{\end{equation}}
\newcommand{\eea}{\end{eqnarray}}

\newcommand{\sla}{\slash \hspace{-0.22cm}}

\def\pslash{p\hspace{-0.18cm}\slash} 
\def\tsigma{\widetilde{\sigma}}

\def\spr{\!\cdot\!}

\def\s#1{{\scriptscriptstyle #1}}

\def\noeq#1{(\ref{#1})}
\def\1eq#1{Eq.~(\ref{#1})}

\def\2eqs#1#2{Eqs.~(\ref{#1}) and~(\ref{#2})}
\def\3eqs#1#2#3{Eqs.~(\ref{#1}),~(\ref{#2}) and~(\ref{#3})}

\def\fig#1{Fig.~\ref{#1}}

\def\diff#1{{\rm d}^#1}
\def\gA{g^2 C_\s A}

\def\ie{{\it i.e.}, }
\def\eg{{\it e.g.}, }

\def\n#1{({\it #1}\,)}

\def\pslash{p\hspace{-0.18cm}\slash}
\def\qslash{q\hspace{-0.19cm}\slash}
\def\kslash{k\hspace{-0.21cm}\slash}
\def\ps#1{p\hspace{-0.18cm}\slash_#1}
\def\tslash{t\hspace{-0.18cm}\slash}

\def\G#1{\Gamma_{#1}}
\def\wG#1{\widehat{\Gamma}_{#1}}

\def\Ga#1#2{\Gamma_{#1}^{\s{#2}}}
\def\wGa#1#2{\widehat{\Gamma}_{#1}^{\s{#2}}}

\def\K#1{K_{#1}}
\def\K#1#2{K_{#1}^{\s{#2}}}
\def\oK#1#2{\overline{K}_#1^{\s{#2}}}

\def\V#1#2{V_{#1#2}^{\s{\rm A}}}

\def\N#1{C_{#1}}

\def\I#1#2{I_{#1}^{#2}}
\def\J#1#2{J_{#1}^{#2}}

\def\d{r(q\spr t)}

\def\R#1{R^{#1}}

\def\Q#1{Q_{#1}}
\def\oQ#1{\overline{Q}_{#1}}

\def\A#1{A_{#1}}
\def\B#1{B_{#1}}

\def\cd{\!\cdot\!}

\def\uE#1{{#1}^\s{\rm{E}}}
\def\dE#1{{#1}_\s{\rm{E}}}

\def\Jgen{{\cal K}}
\def\Igen{\kappa}

\begin{document}

\title{A new method for determining the quark-gluon vertex}

\author{A.~C. Aguilar}
%\email{aguilar@ifi.unicamp.br}
\affiliation{University of Campinas - UNICAMP, 
Institute of Physics ``Gleb Wataghin'' \\
13083-859 Campinas, SP, Brazil}

\author{D. Binosi}
%\email{binosi@ectstar.eu}
\author{D. Iba\~nez}
\affiliation{European Centre for Theoretical Studies in Nuclear
Physics and Related Areas (ECT*) and Fondazione Bruno Kessler, \\Villa Tambosi, Strada delle
Tabarelle 286, 
I-38123 Villazzano (Trento)  Italy}

\author{J. Papavassiliou}
%\email{Joannis.Papavassiliou@uv.es}
\affiliation{\mbox{Department of Theoretical Physics and IFIC, 
University of Valencia and CSIC},
E-46100, Valencia, Spain}

\begin{abstract}

We present a novel non-perturbative approach for
calculating the form factors of the quark-gluon vertex 
in terms of an unknown three-point function, in the Landau gauge.
The key  ingredient  of  this  method  is  the  exact  all-order  relation
connecting the  conventional quark-gluon vertex  with the corresponding
vertex of  the background field  method, which is  Abelian-like.  When
this latter  relation is combined  with the standard  gauge technique,
supplemented by a crucial set of transverse Ward identities, it allows
the approximate  determination of the nonperturbative  behavior of all
twelve form  factors comprising the quark-gluon  vertex, for arbitrary
values of the momenta.  The actual implementation of this procedure is
carried out  in the Landau  gauge, in order  to make contact  with the
results of lattice simulations performed in this particular gauge.  The
most demanding  technical aspect  involves the approximate 
calculation  of the components of the aforementioned 
(fully-dressed) three-point function, using lattice data as
input  for  the  gluon  propagators appearing  in its diagrammatic
expansion.  The  numerical evaluation of the relevant  form factors in
three  special   kinematical  configurations  (soft gluon   and  quark
symmetric  limit,  zero quark  momentum)  is  carried  out in  detail,
finding qualitative agreement with the available  lattice data.  Most
notably, a concrete mechanism  is proposed for explaining the puzzling
divergence  of   one  of  these  form  factors   observed  in  lattice
simulations.

\end{abstract}


\maketitle

\section{Introduction} 

The fundamental vertex that controls the interaction between quarks and gluons 
is considered as  
one of the most important quantities in QCD~\cite{Marciano:1977su,Pascual:1984zb}, and 
a great deal of effort has been devoted to the unraveling of 
its structure and dynamics. 
In fact, its  
nonperturbative properties 
are  essential to a variety of subtle 
mechanisms of paramount theoretical and 
phenomenological relevance.
Indeed, the quark-gluon vertex, which will be denoted by  $\Gamma_{\mu}^{a}(q,r,p)$,
 has a vital impact  
on the dynamics responsible for the breaking of 
chiral symmetry and the subsequent generation of
constituent quark masses~\cite{Maris:2003vk,Roberts:1994dr,Fischer:2003rp,Aguilar:2010cn,Cloet:2013jya}, 
and contributes crucially to the 
formation of the bound states 
that compose the physical spectrum of the theory~\cite{Maris:1999nt,Bender:2002as,Bhagwat:2004hn,Holl:2004qn,Chang:2009zb,Williams:2014iea}.

Despite its physical importance, to date 
the nonperturbative behavior
of this special vertex is still only partially known,
mainly due to a variety of serious technical difficulties\footnote{In perturbation theory, a complete study has been 
carried out at the one-loop level in arbitrary gauges, 
dimensions and kinematics~\cite{Davydychev:2000rt}, whereas  at 
the two- and three-loop order only partial results for specific gauges and kinematics exist~\cite{Chetyrkin:2000fd,Chetyrkin:2000dq}.}.
In particular, its 
rich tensorial structure~\cite{Davydychev:2000rt,Kizilersu:1995iz} leads to a 
considerable proliferation of form factors, which, in addition, depend on three kinematic 
variables (\eg the modulo of two momenta, say $q$ and $r$, and their relative angle). As a result, 
only few (quenched) lattice simulations (in the Landau gauge and 
on modest lattice sizes) 
have been performed~\cite{Skullerud:2002sk,Skullerud:2002ge,Skullerud:2003qu,Skullerud:2004gp,Lin:2005zd,Kizilersu:2006et},
and for a limited number of simple kinematic configurations. 
The situation in the continuum is also particularly cumbersome; indeed, the treatment of this vertex 
in the context of the Schwinger-Dyson equations (SDEs) requires a variety of approximations and truncations~\cite{Bhagwat:2004kj,LlanesEstrada:2004jz,Williams:2014iea,Alkofer:2008tt,Matevosyan:2006bk,Aguilar:2013ac,Rojas:2013tza}, 
and even so, one must deal, at least in principle,  
with an extended system of coupled integral equations (one for each form factor).

There is an additional issue that complicates the extraction of pertinent 
nonperturbative information on the quark-gluon vertex by means of traditional methods, 
which will be of central importance in what follows.
Specifically, in the linear covariant ($R_\xi$) gauges,
$\Gamma_{\mu}^{a}$ satisfies 
a non-linear Slavnov-Taylor identity (STI), imposed by the Becchi-Rouet-Stora-Tyutin (BRST) 
symmetry of the theory. This STI is akin to the 
QED Ward identity (WI) \mbox{$q^{\mu} \Gamma_{\mu}(q,r,p) = S_e^{-1}(r) - S_e^{-1}(p)$}, 
which relates the photon-electron vertex with the electron propagator $S_e$, 
but it is substantially more complicated, because it involves, in addition to the quark propagator $S$, 
contributions from the ghost-sector of the theory (most notably, the so-called ``ghost-quark'' kernel).  
This fact limits considerably the possibility 
of devising a ``gauge technique'' 
inspired Ansatz~\cite{Salam:1963sa,Salam:1964zk,Delbourgo:1977jc,Delbourgo:1977hq} 
for the longitudinal part of $\Gamma_{\mu}^{a}(q,r,p)$. Indeed, whereas 
in an Abelian context the 
longitudinal part of the vertex is expressed exclusively in terms of the Dirac components comprising $S$ 
such that the WI is {\it automatically} satisfied, in the case of the STI  
the corresponding longitudinal part receives   
contributions from additional (poorly known) auxiliary functions and their partial derivatives. 

The applicability of the gauge technique, however, presents an additional difficulty, 
which although intrinsic to this method, acquires its more acute form in a non-Abelian context.
Indeed, as is well-known, the gauge technique leaves the ``transverse'' (automatically conserved) 
part of any vertex (Abelian or non-Abelian) undetermined. 
The amelioration of this shortcoming has 
received considerable attention in the literature, especially 
for the case of the photon-electron vertex, which constitutes the prototype for any type of 
such study~\cite{King:1982mk,Curtis:1990zs,Kizilersu:1995iz,Bashir:1997qt,Kizilersu:2009kg}.
Particularly interesting in this context is the discovery of the so-called 
 ``transverse Ward identities'' (TWIs)~\cite{Takahashi:1985yz,Kondo:1996xn,He:2000we,He:2006my,He:2007zza},
which involve the {\it curl}
of                             the                             vertex,
$\partial_\mu\Gamma_{\nu\cdots}-\partial_\nu\Gamma_{\mu\cdots}$, and can therefore be used, 
at least in principle, to constrain the transverse parts.
The problem is that, unlike WIs, these TWIs  are coupled identities, 
mixing vector and axial terms, and contain non local terms, 
in the form of  gauge-field-dependent  line
integrals~\cite{He:2007zza}. 
However, as was shown in~\cite{Qin:2013mta},
the  induced coupling between  TWIs can in fact be disentangled,
and  the  corresponding  identity  for the  vector  vertex  explicitly
solved.    
Thus   an  Abelian   photon-electron   vertex  satisfying  the
corresponding  WI   and  TWI  could  be  constructed   for  the  first
time~\cite{Qin:2013mta}. 
However, the extension of these results to the non-Abelian sector remains an open issue.
In what follows, for the sake of brevity, we will refer to the framework obtained when the standard  
gauge technique (applied to Abelian WIs)
is supplemented by the  TWIs as the ``improved gauge technique'' (IGT).

The main conclusion of the above considerations is that, whereas the IGT 
constitutes a rather powerful approach for Abelian theories, 
its usefulness for non-Abelian vertices is rather limited. 
It would be clearly most interesting if one could transfer some of the above
techniques to a theory like QCD, and in 
particular, to the quark-gluon vertex. What we propose in the present work is precisely this: 
express the conventional  quark-gluon vertex as a deviation from an 
``Abelian-like'' quark-gluon vertex, 
use the technology derived from the IGT to 
fix this latter vertex, and then compute 
(in an approximate way) the difference between these two vertices.  

The field theoretic framework that enables the  realization of the procedure outlined above is 
the PT-BFM  scheme~\cite{Aguilar:2006gr,Binosi:2007pi,Binosi:2008qk},
which is obtained   through   the    combination   of   the   pinch   technique
(PT)~\cite{Cornwall:1981zr,Cornwall:1989gv,Pilaftsis:1996fh,Binosi:2002ft,Binosi:2003rr,Binosi:2009qm}
with the background field method (BFM)~\cite{Abbott:1980hw}. 
Since within the BFM  the gluon is split
into a  quantum ($Q$) and  a background ($\widehat{A}$)  part, 
two  kinds  of vertices  appear:  vertices  ($\Gamma$)  that have  $Q$
external  lines only  (which correspond to the vertices  appearing  in the
conventional    formulation    of    the    theory)    and    vertices
($\widehat\Gamma$)  that   have  $\widehat{A}$  (or   mixed)  external
lines. Now, interestingly enough, while the former satisfy the usual STIs, the latter 
obey Abelian-like WIs. In addition, a special kind of identities, 
known as 
``background  quantum identities''(BQIs)~\cite{Grassi:1999tp,Binosi:2002ez},
relate the two types of vertices ($\Gamma$ and $\widehat\Gamma$)
 by  means of  auxiliary ghost Green's functions. 
For the specific cases of the quark-gluon vertices the corresponding BQI~\cite{Binosi:2008qk} 
reads schematically [for the detailed dependence on the momenta, see \1eq{BQI}]
\be
\widehat\Gamma_{\mu}=[g^\nu_\mu+\Lambda^\nu_\mu]\Gamma_{\nu} + S^{-1}{K}_\mu + \overline{K}_\mu S^{-1},
\nonumber
\ee
where $\Lambda$ and $K$ are special 
two- and three-point functions, respectively, the origin of which can be ultimately related to the 
antiBRST symmetry of the theory\footnote{The antiBRST symmetry transformations can be obtained from the BRST ones by exchanging the role of the ghost and antighost fields.}.

In the present work, the above BQI will be exploited in order to obtain nontrivial information on all 
twelve form factors of the vertex 
$\Gamma_{\mu}$. Specifically, the main conceptual steps of the approach may be summarized as follows. 
{\bf \n{i}} Since the vertex $\widehat\Gamma_{\mu}$ satisfies a QED-like WI, it will be 
reconstructed using the IGT, following the exact procedure and assumptions (minimal  Ansatz) 
of~\cite{Qin:2013mta}.
{\bf \n{ii}} The two form factors comprising $\Lambda^\nu_\mu$ are known to a high degree of accuracy, 
because they are related to the dressing function of the ghost propagator 
by an exact relation. Since the latter has been obtained in large volume lattice simulations,
as well as computed through SDEs, this part of the calculation is under control.
{\bf \n{iii}} The form of the quark propagator $S$ is obtained from the 
solution of the corresponding quark gap equation.   
{\bf \n{iv}} The functions ${K}_{\mu}$ and $\overline{K}_{\mu}$  
constitute the least known ingredient of this 
entire construction, and must be computed using their diagrammatic expansion, within a 
feasible approximation scheme.
In particular, we employ a version of the ``one-loop dressed'' approximation, where 
the relevant Feynman graphs are evaluated using as input 
fully dressed propagators (obtained from lattice simulations) and bare vertices.

The general procedure outlined above, and developed in the main body of the paper, 
is valid in the context of the linear covariant ($R_\xi$) gauges gauges, for any value of the 
gauge-fixing parameter $\xi$. However, in what follows we will specialize to the particular case of 
$\xi=0$, namely the Landau gauge. The main reason for this choice is the fact that the  
lattice simulations of~\cite{Skullerud:2002sk,Skullerud:2002ge,Skullerud:2003qu,Skullerud:2004gp,Lin:2005zd,Kizilersu:2006et} 
are performed in the Landau gauge; 
therefore, the comparison of our results with the lattice is only possible in this particular gauge. 
An additional advantage of this choice is the fact that the main nonperturbative ingredient entering in our 
diagrammatic calculations, namely the gluon propagator, has been simulated very accurately 
in this gauge~\cite{Bogolubsky:2007ud,Bogolubsky:2009dc},
and will be used as an input (see Sec.~\ref{sec:numres}).

The paper is organized as follows. 
In Sec.~\ref{sec:TF} we set up the theoretical
framework, and review  all the relevant identities (WI,
STI, TWI and BQI) satisfied  by $\Gamma$ and $\widehat\Gamma$.
In Sec.~\ref{sec:complete} we present the main result of our study. 
Specifically, the detailed implementation of the 
procedure outlined above [points {\bf \n{i}}--{\bf \n{iv}}]
furnishes closed expressions for {\it all twelve form factors} comprising  $\Gamma$ 
in a standard tensorial basis, and    
for arbitrary values of the physical momenta. 
Next, in Sec.~\ref{sec:speclim}, we specialize our results to the case of three 
simple kinematic configurations, and derive expressions for the corresponding 
form factors. 
Two of these cases  
(the  ``soft gluon'' and  the  ``symmetric'' limits) 
have already been simulated on the lattice~\cite{Skullerud:2002sk,Skullerud:2002ge,Skullerud:2003qu},
while the third (denominated the ``zero quark momentum'') 
constitutes a genuine prediction of our method. 
 In Sec.~\ref{sec:numres} we carry out the numerical evaluation of the 
expressions derived in the previous section, and then compare with the 
aforementioned lattice results. The coincidence with the lattice results is rather good in most cases. 
In fact, due to the special structure of the 
expressions employed, we are able to suggest a possible mechanism that 
would make one of the ``soft gluon'' form factors diverge at the origin, 
as observed on the lattice; this particular feature has been rather puzzling, and quite resilient to 
a variety of approaches. 
Finally, in  Sec.\ref{sec:concl} we present our discussion and conclusions.
The article ends with two Appendices, where certain technical details are reported.

\section{\label{sec:TF}Theoretical framework}

\begin{figure}[!t]
\begin{center} 
\includegraphics[scale=0.675]{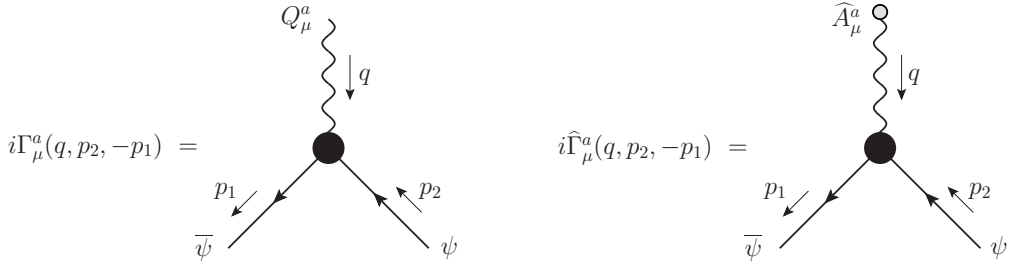}
\caption{\label{quark-vertex}The conventional and background quark-gluon vertex with the momenta routing used throughout the text.}  
\end{center}
\end{figure}

As already mentioned, within the PT-BFM framework one distinguishes  between two quark-gluon vertices, 
depending on the nature of the incoming gluon. Specifically, the vertex formed 
by a quantum gluon (Q) entering into a $ \psi \bar\psi$ pair corresponds to the 
conventional vertex known for the linear renormalizable ($R_\xi$) gauges, 
to be denoted by $\Gamma_\mu^a$; the corresponding three-point function with a 
background gluon ($\widehat{A}$) entering represents instead the PT-BFM vertex and will be denoted by $\widehat{\Gamma}_\mu^a$.  
Choosing the flow of the momenta such that $p_1 = q+p_2$, we then define  (see~\fig{quark-vertex})
\be
i\Gamma_\mu^a(q,p_2,-p_1) = igt^a\Gamma_\mu(q,p_2,-p_1);\qquad 
i\widehat{\Gamma}_\mu^a(q,p_2,-p_1) = igt^a\widehat{\Gamma}_\mu(q,p_2,-p_1),
\ee
where the hermitian and traceless generators $t^{a}$ of the fundamental SU(3) representation are given by $t^{a} = \lambda^{a}/2$, with $\lambda^{a}$ the Gell-Mann  matrices.  Notice that $\Gamma_\mu$ and $\widehat{\Gamma}_\mu$ coincide only at tree-level, where one has $\Gamma^{(0)}_\mu = \widehat{\Gamma}^{(0)}_\mu = \gamma_\mu$.

\subsection{Slavnov-Taylor and (background) Ward identities}

%%%%%%%%%%%%%%%%%%%%%%%%%%%%%%%%%%%%            
%Fig. 1: Quark-gluon vertex
%%%%%%%%%%%%%%%%%%%%%%%%%%%%%%%%%%%
%%%%%%%%%%%%%%%%%%%%%%%%%%%%%%%%%%%
 
One of the most important differences between the two vertices just introduced is that, as a consequence of the background gauge invariance,  $\widehat{\Gamma}_\mu$ obeys a  QED-like WI, instead of the standard STI satisfied by $\Gamma_\mu$~\cite{Abbott:1980hw}. Specifically, one finds
\be
q^\mu \widehat{\Gamma}_{\mu}(q,p_2,-p_1)= S^{-1}(p_1) - S^{-1}(p_2) , 
\label{WI}
\ee
where $S^{-1}(p)$ is  the inverse of the full quark propagator, with
\be
S^{-1}(p) = A(p^2)\,\sla{p} - B(p^2),
\label{qpropAB}
\ee
and $A(p^2)$ and $B(p^2)$ the propagator's Dirac vector and scalar components, respectively. 

%%%%%%%%%%%%%%%%%%%%%%%%%%%%%%%%%%%%            
%Fig. 2: H functions
%%%%%%%%%%%%%%%%%%%%%%%%%%%%%%%%%%%
\begin{figure}[!t]
\begin{center} 
\includegraphics[scale=0.675]{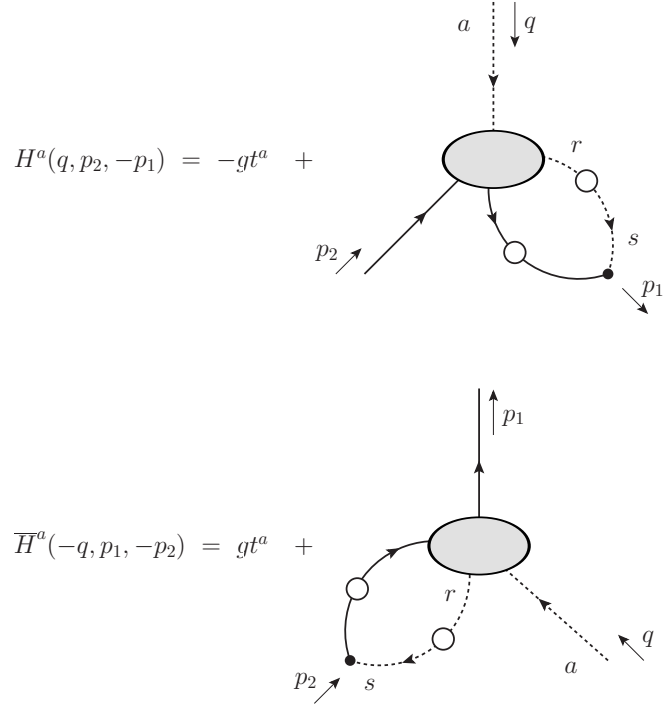}
\caption{\label{H-functions}The ghost kernels $H$ and $\overline{H}$ appearing in the STI satisfied by the quark vertex $\Gamma_\mu$. The composite operators $\psi c^s$ and $\bar\psi c^s$ have the tree-level expressions $-gt^a$ and $gt^a$ respectively.}  
\end{center}
\end{figure}
%%%%%%%%%%%%%%%%%%%%%%%%%%%%%%%%%%%

On the other hand, for the conventional vertex one has
\be
q^\mu\Gamma_{\mu}(q,p_2,-p_1)=F(q^2)\left[S^{-1}(p_1)H(q,p_2,-p_1)-
\overline{H}(-q,p_1,-p_2)S^{-1}(p_2)\right],
\label{STI}
\ee
where $F(q^2)$ denotes the ghost dressing function, which is related to the full ghost propagator $D(q^2)$ through
\be
D(q^2)= \frac{F(q^2)}{q^2},
\ee
whereas the functions $H^a=-gt^aH$ and $\overline{H}^a=gt^a\overline{H}$ correspond to the so-called quark-ghost kernel, and are shown in~\fig{H-functions}. 
It should be stressed that $H$ and $\overline{H}$ are not independent, but are related by ``conjugation''; specifically,  to obtain one from the other, we need to perform the following operations: 
\n{i} exchange $-p_1$ with~$p_2$; 
\n{ii} reverse the sign of all external momenta; 
\n{iii} take the hermitian conjugate of the resulting amplitude. 

Notice that the quark-ghost kernel admits the general decomposition~\cite{Davydychev:2000rt}
\be
H(q,p_2,-p_1)=X_0 \mathbb{I}+X_1 \ps{1}+X_2\ps{2}+X_3\tsigma_{\mu\nu}p^\mu_1p^\nu_2,
\label{HH}
\ee
where $X_i=X_i(q^2,p_2^2,p_1^2)$, and\footnote{ Note the difference between $\tilde\sigma_{\mu\nu}$ and 
the usually defined $\sigma_{\mu\nu} = \frac{i}{2}[\gamma_{\mu},\gamma_{\nu}]$.}
$\tsigma_{\mu\nu}=\frac{1}{2}[\gamma_\mu,\gamma_\nu]$. The decomposition of $\overline{H}$ is then dictated by the aforementioned conjugation operations, yielding
\be
\overline{H}(-q,p_1,-p_2)={\overline{X}}_0 \mathbb{I}+{\overline{X}}_2 \ps{1}
+{\overline{X}}_1\ps{2}+{\overline{X}}_3\tsigma_{\mu\nu}p^\mu_1p^\nu_2,
\label{HH_bar}
\ee
where now $\overline{X_i}={X}_i(q^2,p_1^2,p_2^2)$. At tree-level, one clearly has $ X^{(0)}_0= \overline{X}^{(0)}_0=1$, with the remaining form factors vanishing.

\subsection{Transverse Ward identity}

In addition to the usual WI~\noeq{WI} and STI~\noeq{STI} specifying the \textit{divergence} of the quark-gluon vertex $\partial^\mu\Gamma_\mu$, there exists a set of less familiar identities called transverse Ward identities (TWIs) \cite{Takahashi:1985yz,Kondo:1996xn,He:2000we,He:2006my,He:2007zza,Qin:2013mta} that gives information on the \textit{curl} of the vertex, $\partial_\mu\Gamma_\nu-\partial_\nu\Gamma_\mu$. 
 
Specifically, let us consider the simplified context of an Abelian gauge theory in which a fermion is coupled to a gauge boson through a vector vertex $\Gamma_\mu$ and an axial-vector vertex $\Gamma^\s{\rm A}_\mu$; then the TWIs for these latter vertices read~\cite{Qin:2013mta}
\begin{align}
\label{TWI}
q_\mu \Gamma_\nu(q,p_2,-p_1) - q_\nu \Gamma_\mu(q,p_2,-p_1) &= i[S^{-1}(p_2)\widetilde{\sigma}_{\mu\nu} - \widetilde{\sigma}_{\mu\nu} S^{-1}(p_1)] +2im\Gamma_{\mu\nu}(q,p_2,-p_1)\nonumber \\
&+ t^\lambda \epsilon_{\lambda\mu\nu\rho}\Gamma^\rho_\s{\rm A}(q,p_2,-p_1) + A_{\mu\nu}^\s{\rm V}(q,p_2,-p_1),\nonumber \\
q_\mu \Gamma^\s{\rm A}_\nu(q,p_2,-p_1) - q_\nu \Gamma^\s{\rm A}_\mu(q,p_2,-p_1) &= i[S^{-1}(p_2)\widetilde{\sigma}_{\mu\nu}^5 - \widetilde{\sigma}_{\mu\nu}^5 S^{-1}(p_1)] \nonumber \\
&+ t^\lambda \epsilon_{\lambda\mu\nu\rho}\Gamma^\rho(q,p_2,-p_1) + V_{\mu\nu}^\s{\rm A}(q,p_2,-p_1). 
\end{align}
In the equations above we have set $t=p_1+p_2$ and $\widetilde{\sigma}_{\mu\nu}^5=\gamma_5\widetilde{\sigma}_{\mu\nu}$; 
in addition, $\epsilon_{\lambda\mu\nu\rho}$ is the totally antisymmetric Levi-Civita tensor, while $\Gamma_{\mu\nu}$, $A_{\mu\nu}^\s{\rm V}$, and $V_{\mu\nu}^\s{\rm A}$ represent non-local tensor vertices that appear in this type of identities\footnote{See, {\it e.g.}, \cite{He:2002jg,Pennington:2005mw,He:2006ce} 
for the perturbative one-loop calculations of some of these quantities.}.

As~\1eq{TWI} above shows, the TWIs couple the vector and the axial-vector vertices; however, following the procedure outlined in \cite{Qin:2013mta}, one can disentangle the two vertices, obtaining an identity that involves only one of the two. To do so, let us define the tensorial projectors
\begin{align}
\label{proj}
P_i^{\mu\nu}& = \frac{1}{2}\epsilon^{\alpha\mu\nu\beta}\theta^i_\alpha q_\beta,\qquad i=1,2;& \theta^1_\alpha&=t_\alpha, \quad \theta^2_\alpha=\gamma_\alpha.
\end{align}
Then, due to the antisymmetry of the Levi-Civita tensor, it is easy to realize that 
both tensors annihilate the l.h.s. of the second equation in~\noeq{TWI}; for the vector vertex that we are interested in, one then gets the two identities
\be
[t^\mu\theta^i_\mu q_\rho - (q\spr t)\theta_\rho^i] \Gamma^\rho(q,p_2,-p_1) = P_i^{\mu\nu}\lbrace i[S^{-1}(p_2)\widetilde{\sigma}_{\mu\nu}^5 - \widetilde{\sigma}_{\mu\nu}^5 S^{-1}(p_1)] + V_{\mu\nu}^\s{\rm A}(q,p_2,-p_1)\rbrace, 
\label{transversei}
\ee
which, when used in conjunction with the WI~\noeq{WI}, determine the complete set of form factors characterizing the vertex $\widehat\Gamma_\mu$. 

\subsection{Background-quantum identity}

All the identities described so far (WIs, STIs and TWIs) are the expression at the quantum level of the original BRST symmetry of the SU($N$) Yang-Mills action. However, this action can be also rendered invariant under a less known symmetry that goes under the name of antiBRST~\cite{Curci:1976bt,Ojima:1980da,Baulieu:1981sb}. Then, in~\cite{Binosi:2013cea} it was shown that the requirement that a SU($N$) Yang-Mills action (gauge fixed in an $R_\xi$ gauge) is invariant under both the BRST as well as the corresponding antiBRST symmetry, automatically implies that the theory is quantized in the ($R_\xi$) background field method (BFM) gauge~\cite{Abbott:1980hw}. As an expression of antiBRST invariance, a new set of identities appear, called background quantum identities (BQIs)~\cite{Grassi:1999tp,Binosi:2002ez}, which relate the conventional and PT-BFM vertices.

To obtain the BQI for the quark-gluon vertex in the Landau gauge, 
let us first introduce the auxiliary two-point function 
\bea
\Lambda_{\mu\nu}(q)&=&-i\gA\int_k\!\Delta_\mu^\sigma(k)D(q-k)H_{\nu\sigma}(-q,q-k,k)\nonumber\\
&\equiv&g_{\mu\nu}G(q^2)+\frac{q_\mu q_\nu}{q^2}L(q^2),
\label{Lambda}
\eea
where $C_\s A$ represents the Casimir eigenvalue of the adjoint representation [$C_\s A=N$ for SU(N)], 
$d=4-\epsilon$ is the space-time dimension, and we have introduced the integral measure 
 \mbox{$\int_k=\mu^\epsilon\int\!\diff{d}k/(2\pi)^d$}, 
with $\mu$ the 't Hooft mass. Finally, $H_{\mu\nu}$ is the so called ghost-gluon scattering kernel, and 
$\Delta_{\mu\nu}(q)$ is the gluon propagator, which in the Landau gauge 
reads (see also discussion at the end of this section) 
\be
i\Delta_{\mu\nu}(q)=- i P_{\mu\nu}(q)\Delta(q^2), \qquad  P_{\mu\nu}(q)=g_{\mu\nu}- q_\mu q_\nu/q^2.
\label{Delta}
\ee

In  addition, in this gauge, 
the form factors $G(q^2)$ and  $L(q^2)$ are related to the ghost dressing function $F(q^2)$ by the  all-order relation~\cite{Grassi:2004yq,Aguilar:2009pp}    
\be
F^{-1}(q^2) = 1 + G(q^2) + L(q^2).
\label{funrel}
\ee
Since in four dimensions $L(0)=0$ and $L(q^2)\ll G(q^2)$~\cite{Aguilar:2009pp}, \1eq{funrel} 
is usually replaced by the approximate identity
\be
F^{-1}(q^2) \approx 1 + G(q^2).
\label{funrelapp}
\ee
Notice, however, that, given the subtle nature of the problem at hand, and in order not to distort 
possible cancellations, we will refrain from using \1eq{funrelapp}
in the general derivation of the form factors of $\Gamma_{\mu}$,  employing instead the exact \1eq{funrel}.

%%%%%%%%%%%%%%%%%%%%%%%%%%%%%%%%%%%%            
%Fig. 2: K functions
%%%%%%%%%%%%%%%%%%%%%%%%%%%%%%%%%%%%
\begin{figure}[!t]
\begin{center} 
\includegraphics[scale=0.675]{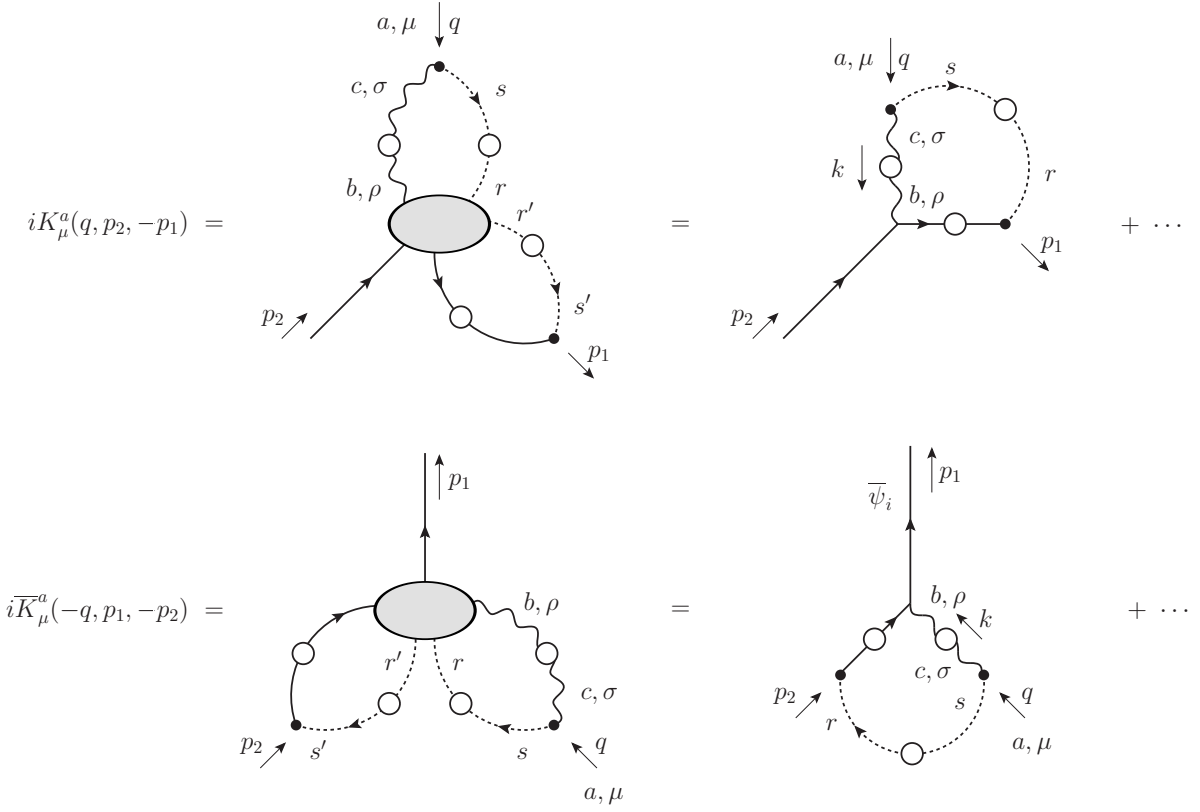}
\caption{\label{fig:K-functions}The auxiliary functions $K_\mu$ and $\overline{K}_\mu$ appearing in the BQI relating the conventional quark vertex $\Gamma_\mu$ with the PT-BFM vertex $\widehat{\Gamma}_\mu$. The composite operator involving $A^c_\sigma \bar c^s$ (with external indices $a,\mu$) has the tree-level expression $gf^{asc}g_{\mu\sigma}$. For later convenience we also show the one-loop dressed approximation of the two functions.}  
\end{center} 
\end{figure}
%%%%%%%%%%%%%%%%%%%%%%%%%%%%%%%%%%%

The BQI of interest (valid in $R_\xi$ gauge) is given by Eq.~(E.13) of the Ref.~\cite{Binosi:2008qk} and reads
\bea
\widehat{\Gamma}_\mu(q,p_2,-p_1)&=&\left[g^\nu_\mu \left(1+G(q^2)\right)+\frac{q_\mu q^\nu}{q^2}L(q^2)
\right]\Gamma_\nu(q,p_2,-p_1)\nonumber \\
&-&S^{-1}(p_1){K}_\mu(q,p_2,-p_1)-\overline{K}_\mu(-q,p_1,-p_2)S^{-1}(p_2),
\label{BQI}
\eea
where the special functions ${K}_\mu$ and $\overline{K}_\mu$ are given in~\fig{fig:K-functions}; 
notice that, as happens for $H$ and $\overline{H}$,  $K$ and $\overline{K}$ are related by conjugation. 
At this point one may appreciate what has been already announced in the introduction, namely that 
the BQI is qualitatively different  
from the WI~\noeq{WI} or the STI~\noeq{STI}, 
since it does not involve the divergence of the vertices. 
Therefore, at least in principle, all form factors of the quark-gluon vertex can be determined by ``solving'' it. 

Let us now contract \1eq{BQI} by $q^\mu$, using simultaneously \2eqs{WI}{STI}, as well as the identity~\noeq{funrel}; it is relatively 
straightforward to establish that the self-consistency of all aforementioned equations imposes an additional relation between the functions  $H$ and~$K$. In Landau gauge one obtains then\footnote{From now on we specialize our procedure to the Landau gauge. Notice, however, that the results can be easily extended to the case of an arbitrary $R_\xi$ gauge by using the aforementioned BRST-antiBRST invariant formulation of the theory~\cite{Binosi:2013cea}. In particular, identities such as  \3eqs{funrel}{HK}{barHbarK} can be generalized to the $\xi\neq0$ case by using the local anti-ghost equation (see Eq.~(4.6) of~\cite{Binosi:2013cea}).}
\be
H(q,p_2,-p_1) = 1 + q^\mu K_\mu(q,p_2,-p_1),
\label{HK}
\ee
as well as the conjugated identity
\be
\overline{H}(-q,p_1,-p_2)=  1 - q^\mu \overline{K}_\mu(-q,p_1,-p_2).
\label{barHbarK}
\ee
\2eqs{HK}{barHbarK} ensure that when the BQI~\noeq{BQI} is contracted with the gluon momentum $q$  it is compatible with both the WI of the background vertex and the STI of the conventional vertex. They are nothing but a consequence of the so-called local antighost equation associated to the antiBRST symmetry~\cite{Binosi:2013cea}.

Coming back to the BQI~\noeq{BQI}, we observe that the term proportional to $L$ triggers the STI~\noeq{STI}; thus, using the relations~\noeq{HK} and~\noeq{barHbarK}, one can write the BQI in its final form in Landau gauge as
\be
{\cal G}(q^2)\Gamma_\mu(q,p_2,-p_1)=\widehat{\Gamma}_\mu(q,p_2,-p_1)+S^{-1}(p_1)\Q{\mu}(q,p_2,-p_1)+\oQ{\mu}(-q,p_1,-p_2)S^{-1}(p_2),
\label{BQI-final}
\ee
where ${\cal G}(q^2)=1+G(q^2)$, and we have defined
\be
\Q{\mu}(q,p_2,-p_1)=K_{\mu}(q,p_2,-p_1)-\frac{q_\mu}{q^2}L(q^2)F(q^2)\left[1+q^\rho K_\rho(q,p_2,-p_1)\right],
\ee
and its conjugated expression
\be
\oQ{\mu}(-q,p_1,-p_2)=\overline{K}_\mu(-q,p_1,-p_2)+\frac{q_\mu}{q^2}L(q^2)F(q^2)\left[1-q^\rho \overline{K}_\rho(-q,p_1,-p_2)\right].
\ee

\begin{table}[!t]
\begin{tabular}{cccc}
\hline
\hline
\bf{Symmetry} & \bf{Associated identities} & \bf{\hspace{0.25cm}Identity type\hspace{0.25cm}} & \bf{Vertices involved} \\
\hline
(Background) gauge & WIs & $\partial^\mu\widehat\Gamma_\mu=\cdots$ & $\widehat\Gamma_\mu$ \\
(Background) gauge & TWIs & $\partial_\mu\widehat\Gamma_\nu-\partial_\nu\widehat\Gamma_\mu=\cdots$ & $\widehat\Gamma_\mu$ \\
BRST & STIs & $\partial^\mu\Gamma_\mu=\cdots$ & $\Gamma_\mu$ \\
antiBRST & BQIs & $\widehat\Gamma_\mu=[g^\nu_\mu+\Lambda^\nu_\mu]\Gamma_\nu+\cdots$ &
$\widehat\Gamma_\mu,\ \Gamma_\mu$ \\
\hline
\hline
\end{tabular}
\caption{\label{tab1}Summary of the SU(N) Yang-Mills symmetries and the associated identities satisfied by the conventional and background quark-gluon vertices.}
\end{table}

The functional identities derived in this section are summarized in Table~\ref{tab1}, together with the symmetries they originate from.

We conclude with a technical issue related to the gluon propagator appearing in \1eq{Delta}. Specifically, 
 in the BFM framework three distinct propagators may be naturally defined:  $\widehat\Delta (q)$, which connects two background gluons 
($\widehat{A} \widehat{A}$), 
$\widetilde\Delta (q)$, which mixes a quantum with a background gluon ($\widehat{A} Q$  and $Q\widehat{A}$), 
and the quantum propagator $\Delta (q)$, which connects two quantum gluons ($QQ$); these three propagators
are related to each other by a set of simple BQIs,  involving only the quantity $\Lambda_{\mu\nu}(q)$~\cite{Grassi:1999tp,Binosi:2002ez}.
There are two important points to remember about $\Delta (q)$. First, $\Delta (q)$ (being the quantum propagator) 
is the only BFM propagator that can propagate inside quantum loops~\cite{Abbott:1980hw}.
Second, $\Delta (q)$ is identical to the conventional gluon propagator of the covariant $R_{\xi}$ gauges 
(and, hence, the same symbol, $\Delta (q)$, is employed), provided of course that the quantum gauge-fixing parameter $\xi_{Q}$
(used in the BFM) is identical to the gauge fixing parameter $\xi$ 
(used in $R_{\xi}$), \ie $\xi_{Q} = \xi$~\cite{Pilaftsis:1996fh,Binosi:2009qm};  
in particular, in the case of the Landau gauge, we have that $\xi_{Q} = \xi=0$. 
This fact, in turn, will permit us later on to use for $\Delta (q)$ 
the corresponding lattice data obtained in the $R_{\xi}$ Landau gauge (see numerical analysis of Sec.~\ref{sec:numres}).

\section{\label{sec:complete}The complete quark gluon vertex}

The next step is to judiciously combine the identities obtained in the previous section, in order to obtain a closed form for the form factors of the quark-gluon vertex $\Gamma_\mu$, in a suitable tensorial basis. Specifically, the procedure to be adopted is detailed in the following: 

\begin{itemize}

\item[\n{i}]We first solve the BQI~\noeq{BQI-final} in order to determine {\it all} the form factors characterizing the conventional vertex (decomposed in a suitable tensorial basis) as a function of the ones appearing in the PT-BFM vertex $\widehat\Gamma_\mu$, and the auxiliary functions $K_\mu$ and $\overline{K}_\mu$ (obviously decomposed in the same basis). As the  BRST and antiBRST symmetries of the PT-BFM formulation guarantee that a solution of the BQI is automatically a solution of the STI, the latter identity is not needed.

\item[\n{ii}]The second step consists in exploiting the Abelian-like nature of $\widehat{\Gamma}_\mu$, 
for determining all of its form factors by simultaneously solving the WI~\noeq{WI} and TWI~\noeq{transversei} it satisfies (IGT). 
The solution obtained will be a function of the propagator components, $A$ and $B$, as well as the non local term $V^\s{\mathrm{A}}$. 

\item[\n{iii}]At this point, one has a formal Ansatz for $\Gamma_\mu$ satisfying all the symmetries of the theory; 
however, the closed form of some of its ingredients is not known. 
To proceed further, we need to make some suitable approximations for the evaluation of  $K$, $\overline{K}$, 
and $V^\s{\mathrm{A}}$. In particular, we will adopt the one-loop dressed approximation for the evaluation of 
the auxiliary functions $K$, $\overline{K}$, whereas for $\widehat{\Gamma}_\mu$ 
we will use the so-called minimal Ansatz of~\cite{Qin:2013mta}, in which the non-local term $V^\s{\mathrm{A}}$ is set directly to zero.
\end{itemize}

In what follows the above main points will be carried out in detail.

\subsection{Tensorial bases}

The procedure outlined above requires 
the definition of a tensorial basis, in order to project out the twelve different components of the vertices and auxiliary functions.
Given the properties of the functions $K_\mu$ and $\overline{K}_\mu$ under conjugation, it is natural to 
employ bases whose components possess simple transformation properties under this operation.

It turns out that there are (at least) two suitable candidates, which we briefly describe below. 

\subsubsection{Transverse/longitudinal basis}

The T+L basis separates the possible contributions to a vector quantity into (four) longitudinal and (eight) transverse form factors. Thus, in the T+L basis all vector quantities  are decomposed according to~\cite{Davydychev:2000rt,Kizilersu:1995iz}
\be
f^\mu(q,p_2,-p_1)=\sum_{i=1}^4f^{\s{L}}_i(q^2,p_2^2,p_1^2)L^\mu_i(q,p_2,-p_1)+\sum_{i=1}^8f^{\s{T}}_i(q^2,p_2^2,p_1^2)T^\mu_i(q,p_2,-p_1),
\ee
where the longitudinal basis vectors read (remember that $t= p_1+p_2$)
\begin{align}
L^\mu_1&=\gamma^\mu;& L^\mu_2&=\tslash t^\mu;& L^\mu_3&=t^\mu;& L^\mu_4&=\widetilde{\sigma}^{\mu\nu}t_\nu;
\label{theLs}
\end{align}
while for the transverse basis vectors we have instead
\begin{align}
T^\mu_1&=p_2^\mu(p_1\cdot q)-p_1^\mu(p_2\cdot q);&
T^\mu_2&=T^\mu_1\tslash;\nonumber\\
T^\mu_3&=q^2\gamma^\mu-q^\mu\qslash;&
T^\mu_4&=T^\mu_1\widetilde{\sigma}_{\nu\lambda}p_1^\nu p_2^\lambda;\nonumber\\
T^\mu_5&=\widetilde{\sigma}^{\mu\nu}q_\nu;&
T^\mu_6&=\gamma^\mu(q\spr t)-t^\mu\qslash;\nonumber\\
T^\mu_7&=-\frac12(q\spr t)L^\mu_4-t^\mu\widetilde{\sigma}_{\nu\lambda}p_1^\nu p_2^\lambda;&
T^\mu_8&=\gamma^{\mu}\widetilde{\sigma}_{\nu\lambda}p_1^\nu p_2^\lambda+p_2^\mu\ps{1}-p_1^\mu\ps{2}.
\label{theTs}\end{align}
It is then relatively straightforward to prove that under conjugation one has the properties
\begin{align}
\overline{L}_i^\mu&= L_i^\mu,\quad i=1,2,3;&
\overline{L}_4^\mu&= -L_4^\mu;\nonumber \\
\overline{T}_i^\mu&= T_i^\mu,\quad i=1,2,3,4,5,7,8;&
\overline{T}_6^\mu&=- T_6^\mu,
\label{conj-prop}
\end{align}
where $L_i^\mu=L_i^\mu(q,p_2,-p_1)$ and $\overline{L}_i^\mu=\overline{L}_i^\mu(-q,p_1,-p_2)$ and similarly for the $T$ tensors.

\subsubsection{Naive conjugated basis}

A second convenient possibility is the NC basis, 
which is obtained by minimally modifying the naive basis of~\cite{Davydychev:2000rt,Kizilersu:1995iz} in order 
to avoid mixing of different tensors under conjugation. 

In this basis the decomposition of a generic vector $f^\mu$ is given by
\be
f^\mu(q,p_2,-p_1)=\sum_{i=1}^{12}f_i(q^2,p_2^2,p_1^2)\N{i}^\mu(q,p_2,-p_1),
\ee
with
\begin{align}
&\N{1}^\mu=\gamma^\mu;&
&\N{2}^\mu=p_2^\mu;&
&\N{3}^\mu=p_1^\mu;&
&\N{4}^\mu=\widetilde{\sigma}^\mu_\nu p_2^\nu;\nonumber \\
&\N{5}^\mu=\widetilde{\sigma}^\mu_\nu p_1^\nu;&
&\N{6}^\mu=p_2^\mu\ps{2};&
&\N{7}^\mu=p_2^\mu\ps{1};&
&\N{8}^\mu=p_1^\mu\ps{2};\nonumber \\
&\N{9}^\mu=p_1^\mu\ps{1};&
&\N{10}^\mu=p_2^\mu\ps{1}\ps{2};&
&\N{11}^\mu=p_1^\mu\ps{1}\ps{2};&
&\N{12}^\mu=\frac12\left(\gamma^{\mu}\ps{1}\ps{2}+\ps{1}\ps{2}\gamma^{\mu}\right).
\end{align}

Then, under conjugation one has the following properties
\begin{align}
\overline{C}_1^\mu&=\N{1}^\mu;&
\overline{C}_2^\mu&=\N{3}^\mu;&
\overline{C}_3^\mu&=\N{2}^\mu;&
\overline{C}_4^\mu&=-\N{5}^\mu;&
\overline{C}_5^\mu&=-\N{4}^\mu;&
\overline{C}_6^\mu&=\N{9}^\mu;\nonumber \\
\overline{C}_7^\mu&=\N{8}^\mu;&
\overline{C}_8^\mu&=\N{7}^\mu;&
\overline{C}_9^\mu&=\N{6}^\mu;&
\overline{C}_{10}^\mu&=\N{11}^\mu;&
\overline{C}_{11}^\mu&=\N{10}^\mu;&
\overline{C}_{12}^\mu&=\N{12}^\mu,
\end{align}
where $\N{i}^\mu=\N{i}^\mu(q,p_2,-p_1)$ and $\overline{C}_i^\mu=\overline{C}_i^\mu(-q,p_1,-p_2)$.

The relations between the form factors in the T+L and NC bases are given in Appendix~\ref{app:NCandT+L}.

%%%%%%%%%%%%%%%%%%%%%%%%%%%%%%%%%%%%%%%%%%%%%%%%%%%%%%%%%%%%%%%%%%%%%%%%%%%%%%%%%%%%%%%%%%%%%%%%%%%%%%%%%%%%%%%%%%%

%%%%%%%%%%%%%%%%%%%%%%%%%%%%%%%%%%%%%%%%%%%%%%%%%%%%%%%%%%%%%%%%%%%%%%%%%%%%%%%%%%%%%%%%%%%%%%%%%%%%%%%%%%%%%%%
\subsection{\label{appIGT} The IGT implementation}

In this subsection we implement the IGT, namely we present 
the general solution of the WI and the TWI, in the T+L basis.

\subsubsection{Ward identity}

For the vertex $\widehat{\Gamma}_\mu$ the ``solution'' of the WI~\noeq{WI} immediately yields for the longitudinal 
form factors~\cite{Ball:1980ay} the expressions 
\begin{align}
\wGa{1}{L}&=\frac{\A{1}+\A{2}}{2};&
\wGa{2}{L}& = \frac{\A{1}- \A{2}}{2(q\spr t)};&  
\wGa{3}{L}&=- \frac{\B{1}-\B{2}}{q\spr t};&
\wGa{4}{L}= 0,
\label{PTBFMlongFF}
\end{align}
where we have defined $A_i=A(p^2_i)$ and $B_i=B(p^2_i)$.
It is then elementary to verify that the resulting ``longitudinal'' vertex satisfies indeed the WI of~\noeq{WI}.

\subsubsection{Transverse Ward identity}

Equations~\noeq{transversei} can be used to determine the remaining (transverse) form factors of $\widehat{\Gamma}_\mu$.
In the T+L basis~\1eq{transversei} yields
\be
(q\spr t)\theta^\mu_i\,\wGa{\mu}{T} = [t^\rho\theta^i_\rho q^\mu - (q\spr t)\theta^\mu_i]\,\wGa{\mu}{L} - iP_i^{\mu\nu}[S^{-1}(p_2)\widetilde{\sigma}_{\mu\nu}^5 - \widetilde{\sigma}_{\mu\nu}^5 S^{-1}(p_1)] - P_i^{\mu\nu}V_{\mu\nu}^\s{\rm A}.
\label{trans1dec}
\ee
Next, introducing the parametrization
\be
P_i^{\mu\nu}V_{\mu\nu}^\s{\rm A} = \V{i}{1} + \V{i}{2}\ps{1} + \V{i}{3}\ps{2} + \V{i}{4} \widetilde{\sigma}_{\mu\nu}p_1^\mu p_2^\nu,
\ee
we obtain for the transverse form factors the general expressions
\begin{align}
%1T
\wGa{1}{T} &= - \frac{1}{2\d}\V{1}{1}, \nonumber \\
%2T
\wGa{2}{T} &= -\frac1{8\d}\left[3(\V{1}{2}+\V{1}{3})-2\V{2}{1}\right], \nonumber \\
%3T
\wGa{3}{T} &= \frac{A_1 - A_2}{2(q\spr t)} + 
\frac1{16\d}
\left\{[3t^2-4(t\spr p_1)]\V{1}{2} + [3t^2-4(t\spr p_2)]\V{1}{3} - 2t^2\V{2}{1}\right\}, \nonumber \\
%4T
\wGa{4}{T} &= \frac1{4r(q\spr t)^2}\left\{2\V{1}{1} - 3(q\spr t)\V{1}{4} - 2(t\spr p_1)\V{2}{2} - 2(t\spr p_2)\V{2}{3}\right\}, \nonumber \\
%5T
\wGa{5}{T} &= -\frac{B_1 - B_2}{q\spr t} - 
\frac1{8\d}\left\{(q\spr t)\V{1}{1} + 2r(\V{1}{4}+\V{2}{2}-\V{2}{3})\right\}, \nonumber \\
%6T
\wGa{6}{T} &= \frac1{16\d}\left\{[4(q\spr p_1) - 3(q\spr t)]\V{1}{2} + [4(q\spr p_2) - 3(q\spr t)]\V{1}{3} + 2(q\spr t)\V{2}{1}\right\}, \nonumber \\
%7T
\wGa{7}{T} &= \frac1{4r(q\spr t)^2}\left\{q^2 \V{1}{1} - 2r(\V{2}{2} + \V{2}{3})\right\}, \nonumber \\
%8T
\wGa{8}{T} &= \frac{A_1 - A_2}{q\spr t} - \frac1{4\d}\left\{(q\spr p_1)\V{1}{2} + (q\spr p_2) \V{1}{3} + r\V{2}{4}\right\},
\label{solTWI}
\end{align}
where we have set $r=r(p_1,p_2)=p_1^2p_2^2 - (p_1\spr p_2)^2$.

By setting all the $\V{i}{j}$ to zero one obtains a minimal Ansatz for the PT-BFM vertex that is compatible with both the WI and the TWI; in this case one finds only three non zero transverse components, namely~\cite{Qin:2013mta}
\begin{align}
\wGa{3}{T}&=\frac{\A{1}- \A{2}}{2(q\spr t)};&
\wGa{5}{T}&=- \frac{\B{1}-\B{2}}{q\spr t};&
\wGa{8}{T}&=-\frac{\A{1}- \A{2}}{q\spr t}.
\end{align}

\subsection{\label{app:generalsol}General solution for arbitrary momenta}

The closed form of the non-Abelian vertex $\Gamma_\mu$, can be finally obtained by solving the BQI~\noeq{BQI-final}. Using the results~\noeq{PTBFMlongFF} we obtain for the longitudinal form factors (T+L basis)
\begin{align}
%1L
{\cal G}_q\Ga{1}{L}&=\left[1-L(q^2)F(q^2)\right]\left\{\wGa{1}{L}+\A{1}\left[\frac12(q\spr t)\K{3}{L}
-(p_1\cd t)\K{4}{L}\right]-\B{1}\K{1}{L}\right. \nonumber\\
&\left.+\A{2}\left[-\frac12(q\spr t)\oK{3}{L}
+(p_2\cd t)\oK{4}{L}\right]-\B{2}\oK{1}{L}\right\},\nonumber\\
%2L
{\cal G}_q\Ga{2}{L}&=\left[1-L(q^2)F(q^2)\right]\left\{\wGa{2}{L}+\A{1}\left[\frac12\K{3}{L}
+\frac{p_1\cd q}{q\spr t}\K{4}{L}\right]-\B{1}\K{2}{L}
\right.\nonumber \\
&\left.+\A{2}\left[\frac12\oK{3}{L}
-\frac{p_2\cd q}{q\spr t}\oK{4}{L}\right]-\B{2}\oK{2}{L}\right\},\nonumber\\
%3L
{\cal G}_q\Ga{3}{L}&=\left[1-L(q^2)F(q^2)\right]\left\{\wGa{3}{L}+\A{1}\left[\frac{p_1\cd q}{q\spr t}\K{1}{L}+(p_1\cd t)\K{2}{L}\right]-\B{1}\K{3}{L}\right.\nonumber\\
&\left.+\A{2}\left[\frac{p_2\cd q}{q\spr t}\oK{1}{L}+(p_2\cd t)\oK{2}{L}\right]-\B{2}\oK{3}{L}\right\},\nonumber\\
%4L
{\cal G}_q\Ga{4}{L}&=\left[1-L(q^2)F(q^2)\right]\left\{\frac{\A{1}}2\left[-\K{1}{L}+(q\spr t)\K{2}{L}\right]-\B{1}\K{4}{L}\right.\nonumber\\
&+\left.\frac{\A{2}}2\left[\oK{1}{L}+(q\spr t)\oK{2}{L}\right]-\B{2}\oK{4}{L}\right\},
\label{general-sol-L}
\end{align}
while, for the transverse form factors we get
\begin{align}
%1T
{\cal G}_q\Ga{1}{T}&=\wGa{1}{T}+\A{1}\left[-\frac1{q\spr t}\K{1}{L}+(p_1\cd t)\K{2}{T}+\K{3}{T}-\K{6}{T}\right]-\B{1}\K{1}{T}\nonumber \\
&+\A{2}\left[\frac1{q\spr t}\oK{1}{L}+(p_2\cd t)\oK{2}{T}+\oK{3}{T}+\oK{6}{T}\right]-\B{2}\oK{1}{T}\nonumber \\
&+\frac2{q^2}L(q^2)F(q^2)\left\{\A{1}\left[\frac{p_1\cd q}{q\spr t}\K{1}{L}+(p_1\spr t)\K{2}{L}\right]-\B{1}\K{3}{L}+\right.\nonumber \\
&+\left.\A{2}\left[\frac{p_2\cd q}{q\spr t}\oK{1}{L}+(p_2\spr t)\oK{2}{L}\right]-\B{2}\oK{3}{L}-\frac{\B{1}-\B{2}}{q\spr t}\right\}
\nonumber \\
%2T
{\cal G}_q\Ga{2}{T}&=\wGa{2}{T}+\A{1}\left[-\frac1{q\spr t}\K{4}{L}+\frac12\K{1}{T}+\frac12(p_1\cd q)\K{4}{T}-\frac12\K{7}{T}\right]-\B{1}\K{2}{T}&\nonumber\\
&+\A{2}\left[-\frac1{q\spr t}\oK{4}{L}+\frac12\oK{1}{T}-\frac12(p_2\cd q)\oK{4}{T}-\frac12\oK{7}{T}\right]-\B{2}\oK{2}{T}\nonumber\\
&+\frac2{q^2}L(q^2)F(q^2)\left\{\A{1}\left[\frac12\K{3}{L}+\frac{p_1\cd q}{q\spr t}\K{4}{L}\right]-\B{1}\K{2}{L}+\right.\nonumber \\
&+\left.\A{2}\left[\frac12\oK{3}{L}-\frac{p_2\cd q}{q\spr t}\oK{4}{L}\right]-\B{2}\oK{2}{L}+\frac{\A{1}-\A{2}}{2(q\spr t)}\right\},
\nonumber
\end{align}
\begin{align}
%3T
{\cal G}_q\Ga{3}{T}&=\wGa{3}{T}+\frac{\A{1}}2\left[\frac12(q\spr t)\K{1}{T}-\frac12(p_1\cd t)(q\spr t)
\K{4}{T}-\K{5}{T}\right]-\B{1}\K{3}{T}&\nonumber\\
&+\frac{\A{2}}2\left[-\frac12(q\spr t)\oK{1}{T}+\frac12 (p_2\cd t)(q\spr t)
\oK{4}{T}-\oK{5}{T}\right]-\B{2}\oK{3}{T}
\nonumber\\
&+\frac1{q^2}L(q^2)F(q^2)\left\{\A{1}\left[\frac12(q\spr t)\K{3}{L}-(p_1\spr t)\K{4}{L}\right]-\B{1}\K{1}{L}
\right.\nonumber\\
&+\left.\A{2}\left[-\frac12(q\spr t)\oK{3}{L}+(p_2\spr t)\oK{4}{L}\right]-\B{2}\oK{1}{L}+\frac12(\A{1}+\A{2})
\right\},\nonumber \\
%4T
{\cal G}_q\Ga{4}{T}&=\wGa{4}{T}+\A{1}\left[\K{2}{T}-\frac{2}{q\spr t}\K{3}{T}+\frac1{q\spr t}\K{8}{T}\right]-\B{1}\K{4}{T}+\A{2}\left[\oK{2}{T}+\frac{2}{q\spr t}\oK{3}{T}-\frac1{q\spr t}\oK{8}{T}\right]-\B{2}\oK{4}{T}\nonumber\\
&+\frac4{q^2(q\spr t)}L(q^2)F(q^2)\left\{\frac{\A{1}}2\left[-\K{1}{L}+(q\spr t)\K{2}{L}\right]-\B{1}\K{4}{L}+\frac{\A{2}}2\left[\oK{1}{L}+(q\spr t)\oK{2}{L}\right]-\B{2}\oK{4}{L}\right\},\nonumber \\
%5T
{\cal G}_q\Ga{5}{T}&=\wGa{5}{T}+\frac{\A{1}}2\left[-\K{1}{L}-q^2\K{3}{T}-(q\spr t)\K{6}{T}-(p_1\cd t)\K{8}{T}\right]-\B{1}\K{5}{T}\nonumber\\
&+\frac{\A{2}}2\left[-\oK{1}{L}-q^2\oK{3}{T}-(q\spr t)\oK{6}{T}-(p_2\cd t)\oK{8}{T}\right]-\B{2}\oK{5}{T},\nonumber\\
%6T
{\cal G}_q\Ga{6}{T}&=\wGa{6}{T}+\frac{\A{1}}2\left[-\K{3}{L}-\frac{q^2}2\K{1}{T}+\frac{q^2}2(p_1\cd t)\K{4}{T}-\K{5}{T}-(p_1\cd t)\K{7}{T}\right]-\B{1}\K{6}{T}&\nonumber\\
&+\frac{\A{2}}2\left[\oK{3}{L}+\frac{q^2}2\oK{1}{T}-\frac{q^2}2(p_2\cd t)\oK{4}{T}+\oK{5}{T}+(p_2\cd t)\oK{7}{T}\right]-\B{2}\oK{6}{T},\nonumber\\
%7T
{\cal G}_q\Ga{7}{T}&=\wGa{7}{T}+\A{1}\left[-\K{2}{L}-\frac{q^2}{q\spr t}\K{3}{T}-\K{6}{T}+\frac{p_1\cd q}{q\spr t}\K{8}{T}\right]-\B{1}\K{7}{T}&\nonumber\\
&+\A{2}\left[-\oK{2}{L}+\frac{q^2}{q\spr t}\oK{3}{T}+\oK{6}{T}+\frac{p_2\cd q}{q\spr t}\oK{8}{T}\right]-\B{2}\oK{7}{T}\nonumber\\
&+\frac2{q\spr t}L(q^2)F(q^2)\left\{\frac{\A{1}}2\left[-\K{1}{L}+(q\spr t)\K{2}{L}\right]-\B{1}\K{4}{L}+\frac{\A{2}}2\left[\oK{1}{L}+(q\spr t)\oK{2}{L}\right]-\B{2}\oK{4}{L}
\right\},\nonumber \\
%8T
{\cal G}_q\Ga{8}{T}&=\wGa{8}{T}+\A{1}\left[\K{4}{L}-\K{5}{T}+\frac12(q\spr t)\K{7}{T}\right]-\B{1}\K{8}{T}+\A{2}\left[-\oK{4}{L}-\oK{5}{T}-\frac12(q\spr t)\oK{7}{T}\right]-\B{2}\oK{8}{T}.
\label{general-sol-T}
\end{align}
In the formulas above, 
\begin{align}
{\cal G}_q&=1+G(q^2);&
\K{i}{T,L}&=\K{i}{T,L}(q^2,p_2^2,p_1^2);&
\oK{i}{T,L}&=\oK{i}{T,L}(q^2,p_1^2,p_2^2).&
\end{align}

As far as the longitudinal terms are concerned,
it should be noticed that 
the form of~\1eq{general-sol-L} is dictated by the required compatibility between the STI and the BQI. Indeed, using~\1eq{funrel} we get the relation $1-L(q^2)F(q^2)={\cal G}_qF(q^2)$, so that the ${\cal G}_q$ simplifies and one is left with the result we would have obtained starting directly from the STI~\noeq{STI} after using~\2eqs{HK}{barHbarK} to trade the form factors appearing in the $H$ and $\overline H$  for the ones appearing in $K$ and $\overline K$. This is not the case for the transverse form factors, where indeed no such pattern is found.

\section{\label{sec:speclim} Some special kinematic limits}

Here we specialize the general solution reported in the previous section 
to the two kinematic configurations that have been simulated on the lattice~\cite{Skullerud:2002sk,Skullerud:2002ge,Skullerud:2003qu}, 
corresponding to the soft gluon limit $p_1\to p_2$ (or $q\to0$) and the symmetric limit $p_1\to-p_2$. 
A third interesting limit in which the quark momenta $p_2$ is set to zero, will be also discussed.

\subsection{Soft-gluon limit}

The solution of the BQI in this limit can be obtained by letting   $p_1\to p_2$ in the general solution presented in~\2eqs{general-sol-L}{general-sol-T}. Although several of the expressions appearing there seem singular in this limit, 
it should be noticed that this is not the case. The reason is that 
whenever $p_1\to\pm p_2$,  the form factors $\oK{i}{L,T}$ and $\K{i}{L,T}$ also coincide (up to a sign); indeed, the conjugation properties~\noeq{conj-prop} gives the relations
\begin{align}
\oK{i}{L}(q^2,p_1^2,p_2^2)&=\K{i}{L}(q^2,p_1^2,p_2^2)& i&=1,2,3;&\oK{4}{L}(q^2,p_1^2,p_2^2)&=-\K{4}{L}(q^2,p_1^2,p_2^2)\nonumber \\
\oK{i}{T}(q^2,p_1^2,p_2^2)&=\K{i}{T}(q^2,p_1^2,p_2^2)& i&=1,2,3,4,5,7,8;&
\oK{6}{T}(q^2,p_1^2,p_2^2)&=-\K{6}{L}(q^2,p_1^2,p_2^2).
\label{soft-symm-rels}
\end{align}

As a result, all potentially divergent terms cancel out and one is left with a well defined result.
In particular, since the limit $p_1\to p_2$  also implies that $q\to0$, all the transverse tensor structures~\noeq{theTs} vanish identically. 
The vertex is therefore purely longitudinal, and after setting $p_1=p_2=p$, one finds that the $L_{i}^{\mu}$ vectors reduce to
\begin{align}
L_1^\mu&=\gamma^\mu;&
L_2^\mu&=4\pslash p^\mu;&
L_3^\mu&=2p^\mu;&
L_4^\mu&=2\widetilde{\sigma}^{\mu\nu}p_\nu.
\end{align}
Redefining the basis vectors so that they are simply given by~$\{\gamma^\mu,\pslash p^\mu,p^\mu,\widetilde{\sigma}^{\mu\nu}p_\nu\}$ with corresponding form factors~$\{\Ga{1}{},\Ga{2}{},\Ga{3}{},\Ga{4}{}\}$ and $\{\K{1}{},\K{2}{},\K{3}{},\K{4}{}\}$, one obtains the results 
\begin{align}
F_0^{-1}\G{1}&=A\left(1-2p^2\K{4}{}\right)-2B\K{1}{},\nonumber\\
F_0^{-1}\G{2}&=2A'+2A\left(\K{3}{}+\K{4}{}\right)-2B\K{2}{},\nonumber \\
F_0^{-1}\G{3}&=-2B'+2A\left(\K{1}{}+p^2\K{2}{}\right)-2B\K{3}{},\nonumber \\
\G{4}&=0,
\label{softlimit-results-Mink}
\end{align}   
where $F^{-1}_0=F^{-1}(0)$, $A=A(p^2)$, $B=B(p^2)$, $\K{i}{}=\K{i}{}(p^2)$, and a prime denotes derivative with respect to $p^2$. 

We conclude this subsection by noticing that in the soft gluon limit the identities~\noeq{HK} and \noeq{barHbarK} yield an {\it all-order} constraint on the form of $H$ and $\overline{H}$. To see this, let us observe that the Taylor expansion of a function $f(q,p_2,-p_1)$ when $q\to0$, and $p_1=p_2=p$ reads
\be
f(q,p_2,-p_1) = f(0,p,-p) + q^\mu \left. \frac{\partial}{\partial q^\mu}f(q,p_2,-p_1)\right\vert_{q=0} + {\cal O}(q^2),
\label{generalTayl}
\ee
where the (possible) Lorentz structure of the function $f$ has been suppressed.  Specializing this result to the identities~\noeq{HK}, one obtains the (all-order) conditions 
\be
H(0,p,-p)=1\qquad\Longrightarrow \qquad X_1(0,p^2,p^2)=-X_2(0,p^2,p^2); \quad X_0(0,p^2,p^2)=1,
\label{softX}
\ee
where we have used the form factor decomposition of~\1eq{HH}.  Clearly, an equivalent result holds for $\overline{H}$ and its corresponding form factors.

\subsection{Symmetric limit}

The symmetric limit, in which  $p_1\to-p_2$, is subtler than the previous case. 
The relations listed in \1eq{soft-symm-rels} remain valid also in this limit, 
thus leading to a finite result for the expressions~\noeq{general-sol-L} 
and~\noeq{general-sol-T}; nevertheless, one finds that only one longitudinal basis tensor~\noeq{theLs} and two transverse tensors~\noeq{theTs} survive in this limit, namely
\begin{align}
L_1^\mu&=\gamma^\mu;&
T_3^\mu&=4\left(p^2\gamma^\mu-p^\mu\pslash\right);&
T_5^\mu&=-2\widetilde{\sigma}^{\mu\nu}p_\nu.
\end{align}
However, as in the previous case, there are in principle four independent tensors in the basis: we are clearly missing $p^\mu$. 

Thus, we arrive to the conclusion that in the T+L basis the symmetric limit is singular, and one cannot get the results by taking directly this limit  in the general solution~\noeq{general-sol-L} and~\noeq{general-sol-T}. The way to proceed is instead the following: \n{i} first, use the relations~\noeq{T+LtoNaiveconj} before taking any limit to get the general solution in the naive conjugated basis; \n{ii} next, take the symmetric limit of this solution, given that this basis is well behaved in this limit, giving rise to the four independent tensors needed;
\n{iii} go back to the T+L basis using~\2eqs{NaiveconjtoT+L:L}{NaiveconjtoT+L:T}.

Following this procedure, and redefining the basis vectors to be, as in the previous limit,~$\{\gamma^\mu,\pslash p^\mu,p^\mu,\widetilde{\sigma}^{\mu\nu}p_\nu\}$, one obtains the results
\begin{align}
{\cal G}_{2p}\G{1}&=\wG{1}+2p^2A\K{4}{}-2B\K{1}{},\nonumber\\
{\cal G}_{2p}\G{2}&=\wG{2}-2A\left(\K{3}{}+\K{4}{}\right)-2B\K{2}{}-\frac1{p^2}L_{2p}F_{2p}\left[A\left(1-2p^2\K{3}{}\right)-2B\left(\K{1}{}+p^2\K{2}{}\right)\right]
\nonumber\\
{\cal G}_{2p}\G{3}&=\wG{3},\nonumber \\
{\cal G}_{2p}\G{4}&=\wG{4}+2A\K{1}{}-2B\K{4}{},
\end{align}
where $A=A(p^2)$, $B=B(p^2)$ and
\begin{align}
{\cal G}_{2p}&=1+G(4p^2);&
F_{2p}&=F(4p^2);&
L_{2p}&=L(4p^2).
\end{align}

Within this basis, the solution of the WI and TWI~\noeq{PTBFMlongFF} and~\noeq{solTWI} gives the relations 
\be
\wG{1}+p^2\wG{2}=A;\qquad \wG{3}=0;\qquad \wG{4}=2B'
\ee
and therefore one gets the final results 
\begin{align}
F^{-1}_{2p}\left(\G{1}+p^2\G{2}\right)&=A\left(1-2p^2\K{3}{}\right)-2B\left(\K{1}{}+p^2\K{2}{}\right),\nonumber \\
\G{3}&=0,\nonumber\\
{\cal G}_{2p}\G{4}&=2B'+2A\K{1}{}-2B\K{4}{}.
\end{align}

As the above results clearly show, in the naive conjugated basis it is not possible to disentangle the form factors $\G{1}$ and $\G{2}$. This, however, can be achieved by going back to the T+L basis $\{\gamma^\mu,p^\mu,p^2\gamma^\mu-\pslash p^\mu,\widetilde{\sigma}^{\mu\nu}p_\nu\}$ in which the corresponding form factors~$\{\Ga{1}{L},\Ga{3}{L},\Ga{3}{T},\Ga{5}{T}\}$ can be obtained from the previous ones through the relations
\begin{align}
\Ga{1}{L}&=\G{1}+p^2\G{2};& \Ga{3}{L}&=\G{3};& \Ga{3}{T}&=-\G{2};&\Ga{5}{T}&=\G{4},
\end{align}
and similarly for $\{\K{1}{L},\K{3}{L},\K{3}{T},\K{5}{T}\}$; one then obtains
\footnote{We notice that the terms proportional to $\K{3}{L}$ are precisely those 
that one would miss by taking directly the symmetric limit of the T+L solution~\noeq{general-sol-L} and~\noeq{general-sol-T}.} 
\begin{align}
F_{2p}^{-1}\Ga{1}{L}&=A-2p^2A\K{3}{L}-2B\K{1}{L},\nonumber\\
\Ga{3}{L}&=0,\nonumber\\
{\cal G}_{2p}\Ga{3}{T}&=2A'+2A\left[\left(1-L_{2p}F_{2p}\right)\K{3}{L}+\K{5}{T}+\frac1{2p^2}L_{2p}F_{2p}\right]-2B\left[\K{3}{T}+\frac1{p^2}L_{2p}F_{2p}\K{1}{L}\right],\nonumber \\
{\cal G}_{2p}\Ga{5}{T}&=2B'+2A\left(\K{1}{L}+p^2\K{3}{T}\right)-2B\K{5}{T}.
\label{symlimit-results-M} 
\end{align}

However, on the lattice in the Landau gauge and for a momentum configuration other than the soft gluon, what they have measured is only the combination~$P^\nu_\mu\Gamma_\nu$, that is
\begin{equation}
P^\nu_\mu(p)\Gamma_\nu=P^\nu_\mu(p)\gamma_\nu(
\Ga{1}{L}+p^2\Ga{3}{T})+\widetilde{\sigma}_{\mu\nu}p^\nu\Ga{5}{T},
\label{prjvert}
\end{equation}
yielding
\begin{align}
{\cal G}_{2p}(\Ga{1}{L}+p^2\Ga{3}{T})&=2p^2A'+A(1+2p^2\K{5}{T})-2B(\K{1}{L}+p^2\K{3}{T}),\nonumber \\
{\cal G}_{2p}\Ga{5}{T}&=2B'+2A\left(\K{1}{L}+p^2\K{3}{T}\right)-2B\K{5}{T}.
\label{symmff}
\end{align}
Evidently, the multiplication of  $\Ga{3}{T}$ by $p^2$ removes the potentially 
IR divergent terms; thus, one expects the corresponding form factor measured on the lattice to be finite.

\subsection{Zero quark momentum}

We now set to zero the quark momentum $p_2$, so that $q=p_1=p$. This limit is well defined in any of the two bases introduced earlier, and the corresponding form factors can be obtained directly form our general solution~\noeq{general-sol-L} and~\noeq{general-sol-T}. However, the form factors $\K{i}{L,T}(p^2,0,p^2)$ and $\oK{i}{L,T}(p^2,p^2,0)$ do not coincide anymore, and need to be evaluated separately.  
Defining the basis tensors to be $\{\gamma^\mu,p^\mu,p^2\gamma^\mu-\pslash p^\mu,\widetilde{\sigma}^{\mu\nu}p_\nu\}$ with the corresponding form factors~$\{\Ga{1}{L},\Ga{3}{L},\Ga{3}{T},\Ga{5}{T}\}$, $\{\K{1}{L},\K{3}{L},\K{3}{T},\K{5}{T}\}$, and $\{\oK{1}{L},\oK{3}{L},\oK{3}{T},\oK{5}{T}\}$, we obtain the following results
\begin{align} 
F^{-1} \Ga{1}{L} &= A(1 + p^2\K{3}{L}) - B\K{1}{L} - B_0\oK{1}{L}, \nonumber \\
F^{-1} \Ga{3}{L} &= -\frac{1}{p^2}(B-B_0) + A\K{1}{L} - B\K{3}{L} - B_0\oK{3}{L}, \nonumber \\
{\cal G}\Ga{3}{T} &=- A(\K{3}{L} + \K{5}{T}) - B\K{3}{T} - B_0\oK{3}{T} + \frac{1}{p^2}L_p F_p\left[A(1+p^2\K{3}{L}) - B\K{1}{L} - B_0\oK{1}{L}\right], \nonumber \\
{\cal G}\Ga{5}{T} &=  -\frac{1}{p^2}(B-B_0) - A(\K{1}{L}+p^2\K{3}{T}) - B\K{5}{T} - B_0\oK{5}{T},
\label{solBQIonep}
\end{align}
with the usual definitions $A=A(p^2)$,  $B=B(p^2)$, as well as $B_0=B(0)$.

On the lattice one focuses on the projected vertex~\noeq{prjvert}, for which one has the two form factors
\begin{align}
{\cal G}(\Ga{1}{L}+p^2\Ga{3}{T})&=A(1-p^2\K{5}{T})-B(\K{1}{L}+p^2\K{3}{T})-B_0(\oK{1}{L}+p^2\oK{3}{T}),\nonumber \\
{\cal G}\Ga{5}{T}&=-\frac{1}{p^2}(B-B_0) - A(\K{1}{L}+p^2\K{3}{T}) - B\K{5}{T} - B_0\oK{5}{T}.
\label{ffzero}
\end{align}

\section{\label{sec:numres}Numerical results and comparison with lattice data}

In this section we carry out a numerical study of the form factors of the quark-gluon vertex 
in the various kinematical limits studied in the previous section.

\subsection{The one-loop dressed approximation for the auxiliary functions}

As a first step in our numerical study, we need to identify a suitable approximation for the functions $K_\mu$ and $\overline{K}_\mu$ in order to determine the corresponding form factors $\K{i}{}$ and $\oK{i}{}$, which ultimately characterize the quark-gluon vertex. 
In what follows we will use the one-loop dressed approximation (see~\fig{fig:K-functions}), in which the propagators are fully dressed while vertices are retained at tree-level (see \fig{fig:K-functions} again). This yields the following expressions 
\begin{align}
K_\mu(q,p_2,-p_1)&=\frac i2g^2C_A\int_k\!S(k+p_2)\gamma^\nu P_{\mu\nu}(k)\Delta(k^2)D(k-q),\nonumber \\
\overline{K}_\mu(-q,p_1,-p_2)&=\frac i2g^2C_A\int_k\!\gamma^\nu S(p_1-k) P_{\mu\nu}(k)\Delta(k^2)D(k-q).
\label{kkk}
\end{align}
%As the two functions are related by conjugation, in addition we can concentrate on the calculation of $K_\mu$ only.  

It turns out that the best and most expeditious strategy for projecting out the various components of this function is to use the naive conjugated basis, eventually passing to the T+L basis using the formulas~\noeq{NaiveconjtoT+L:L} and~\noeq{NaiveconjtoT+L:T}. For general values of the $p_i$ momenta the calculation is carried out in Appendix~\ref{1ldints}; here we will study the
limiting cases singled out in the previous section (notice that, in the case of the soft gluon and symmetric limit, one cannot obtain the corresponding results as a direct limit of the general results).

\subsubsection{Soft-gluon and symmetric limit}

In the limit  $p_1\to\pm p_2$ one can concentrate on the calculation of $K_\mu$ only, as in this case $K$ and $\overline{K}$ coincide. Thus, we start by writing
\be
K_\mu(p)=\frac i2g^2C_A\int_k\left(\kslash+\pslash\right)\gamma^\nu P_{\mu\nu}(k)R^\s{A}(k,p)+\frac i2g^2C_A\int_k\gamma^\nu P_{\mu\nu}(k)R^\s{B}(k,p),
\ee
where we have defined
\be
\R{f}(k,p)=\frac{f(k+p)\Delta(k^2)}{A^2(k+p)(k+p)^2-B^2(k+p)}{\cal D}(k,p),
\label{Rf-soft-symm}
\ee
and
\be
{\cal D}(k,p)=\left\{
\begin{array}{ll}
D(k), &\quad \rm{soft\ gluon\ limit;}\\
D(k+2p),&\quad \rm{quark\ symmetric\ limit.}
\end{array}
\right.
\ee

We next introduce the integrals
\begin{align}
\I{0}{f}(p)&=\frac i2g^2C_A\int_k\! R^f(k,p),\nonumber \\
\I{\mu}{f}(p)&=\frac i2g^2C_A\int_k\!k_\mu R^f(k,p)=\I{1}{f}(p^2)p_\mu,\nonumber \\
\I{\mu\nu}{f}(p)&=\frac i2g^2C_A\int_k\!\frac{k_\mu k_\nu}{k^2}R^f(k,p)=\J{1}{f}(p^2)g_{\mu\nu}+\J{2}{f}(p^2)p_\mu p_\nu,
\end{align}
with, correspondingly,
\begin{align}
\I{1}{f}(p^2)&=\frac{p^\mu}{p^2}\I{\mu}{f}(p);&
\J{1}{f}(p^2)&=\frac13P^{\mu\nu}(p)\I{\mu\nu}{f}(p);&
\J{2}{f}(p^2)&=\frac1{3p^2}\left(4\frac{p^\mu p^\nu}{p^2}-g^{\mu\nu}\right)\I{\mu\nu}{f}(p).
\end{align}
Notice that not all these form factors are independent, since one has the constraint
\be
4\J{1}{f}(p^2)=\I{0}{f}(p^2)-p^2\J{2}{f}(p^2).
\ee

Writing finally
\be
K_\mu(p)=\gamma_\mu\K{1}{}(p^2)+\pslash p_\mu \K{2}{}(p^2)+p_\mu \K{3}{}(p^2)+\widetilde{\sigma}_{\mu\nu}p^\nu\K{4}{}(p^2),
\ee
we obtain the results
\begin{align}
\K{1}{}(p^2)&=\I{0}{\s{B}}(p^2)-\J{1}{\s{B}}(p^2)=\frac i6g^2C_A\int_k\!\left[2+\frac{(k\!\cdot\!p)^2}{k^2p^2}\right]R^\s{B}(k,p),
\nonumber\\
\K{2}{}(p^2)&=-\J{2}{\s{B}}(p^2)=\frac i{6p^2}g^2C_A\int_k\!\left[1-4\frac{(k\!\cdot\!p)^2}{k^2p^2}\right]R^\s{B}(k,p),\nonumber \\
\K{3}{}(p^2)&=%\I{0}{\s{A}}(p^2)-\J{1}{\s{A}}(p^2)-p^2\J{2}{\s{A}}(p), 
3\J{1}{\s{A}}(p^2)=\frac i2g^2C_A\int_k\!\left[1-\frac{(k\!\cdot\!p)^2}{k^2p^2}\right]R^\s{A}(k,p),
\nonumber \\
\K{4}{}(p^2)&=-\I{0}{\s{A}}(p^2)-\I{1}{\s{A}}(p^2)+\J{1}{\s{A}}(p^2)=-\frac i6g^2C_A\int_k\!\left[2+3\frac{(k\spr p)}{p^2}+\frac{(k\!\cdot\!p)^2}{k^2p^2}\right]R^\s{A}(k,p).
\label{theKs}
\end{align}
Notice the $1/p^2$ factor multiplying the $K_2(p^2)$ function; we will return to this important point shortly.

\subsubsection{Zero quark momentum}

In this case one has to consider both $K$ and $\overline{K}$, as when $p_2=0$ the two functions do not coincide. For $K_\mu$, after defining
\be
\R{f}(k,p)=\frac{f(k^2)\Delta(k^2)D(k+p)}{A^2(k^2)k^2-B^2(k^2)},
\label{Reqzq}
\ee  
one finds that $\K{3}{}(p^2)=0$, $\K{1}{}$ and $\K{2}{}$ are given in~\1eq{theKs} with $R^\s{B}$ obtained from~\1eq{Reqzq} above, and finally 
\begin{equation}
\K{4}{}(p^2)=\I{1}{\s{A}}(p^2)=\frac i2g^2C_A\int_k\!\frac{(k\!\cdot\!p)}{p^2}R^\s{A}(k,p).
\end{equation}

For $\overline{K}_\mu$ one has instead
\begin{equation}
\R{f}(k,p)=\frac{f(k+p)\Delta(k^2)D(k+p)}{A^2(k+p)(k+p)^2-B^2(k+p)},
\label{Reqzq1}
\end{equation}
and one gets for the $\oK{i}{}$ the corresponding results of~\1eq{theKs} for $\K{i}{}$, in which $R^f$ is replaced by the expression above and $\K{4}{}$ gets an extra minus sign.

\subsection{Passing to the Euclidean space}

In order to pass from Minkowskian to Euclidean space, let us define
\begin{align}
\gamma^0&\to\uE{\gamma}_4;&
\gamma^j&\to i\uE{\gamma}_j;&
k^0&\to i\uE{k}_4;&
k^j\to-\uE{k}_j.
\end{align}
Then, with the signature of the Minkowski metric being $(+,-,-,-)$, one has the replacement rules 
\begin{align}
{\rm d}^4k&\to i{\rm d}^4\dE{k};& 
\kslash&\to i\dE{\kslash};& 
k\cdot q&\to -\dE{k}\cdot\dE{q};& 
k^2&\to-\dE{k}^2. 
\end{align}
On the one hand, these rules are enough to convert to their  Euclidean counterparts scalar expressions; specifically one has 
\begin{align}
&A_\s{\rm{E}}(p^2_\s{\rm{E}})=A(-p^2);&
&B_\s{\rm{E}}(p^2_\s{\rm{E}})=B(-p^2);\nonumber \\
&F_\s{\rm{E}}(p^2_\s{\rm{E}})=F(-p^2);&
&\Delta_\s{\rm{E}}(p^2_\s{\rm{E}})=-\Delta(-p^2);\nonumber\\
&K_{1,3,4}^\s{\rm{E}}(p^2_\s{\rm{E}})=K_{1,3,4}(-p^2);&
&K_2^\s{\rm{E}}(p^2_\s{\rm{E}})=-K_2(-p^2);\nonumber \\
&\overline{K}_{1,3,4}^{\s{\rm{E}}}(p^2_\s{\rm{E}})=\overline{K}_{1,3,4}(-p^2);&
&\overline{K}_2^\s{\rm{E}}(p^2_\s{\rm{E}})=-\overline{K}_2(-p^2) \,.
\end{align}
However,  they are not sufficient to specify how to proceed in the case of 
a four-vector quantity like the quark-gluon vertex; to accomplish the conversion, we follow the prescription of~\cite{Skullerud:2003qu}.
Specifically,  
first we form a Minkowski scalar 
by contracting $\Gamma_{\mu}$ with $\gamma^\mu$, and then we demand that the resulting expression  
be identical to the one obtained if 
we had started directly from the Euclidean expression, and had assumed that all the Euclidean 
form factors are equal to the corresponding  Minkowski ones evaluated at negative momenta, 
\begin{equation}
\Gamma_i(\dE{q}^2,p_{2\s{\rm E}}^2,p_{1\s{\rm E}}^2)=\Gamma_i(-q^2,-p_2^2,-p_1^2).
\end{equation}
In the kinematic configurations of interest, which involves only one momentum scale $p$, this prescription yields the NC tensor basis $\{\uE{\gamma}_\mu,i\uE{p}_\mu,-\pslash_\s{\rm{E}} \uE{p}_\mu,i\uE{\widetilde{\sigma}}_{\mu\nu}\uE{p}_\nu\}$ or the T+L basis 
$\{\uE{\gamma}_\mu,i\uE{p}_\mu,-\dE{p}^2\uE{\gamma}_\mu+\pslash_\s{\rm{E}} \uE{p}_\mu,i\uE{\widetilde{\sigma}}_{\mu\nu}\uE{p}_\nu\}$. 

Finally, integrals will be performed using the following spherical coordinates:
\begin{align}
& x=p^2;\qquad y=k^2;\qquad z=(k + p)^2=x+y+2\sqrt{xy}\cos\theta;\nonumber\\
& \int_{k_\s{\mathrm{E}}}=\frac1{(2\pi)^3}\int_0^\pi\!\mathrm{d}\theta\,\sin^2\theta\int_0^\infty\!\mathrm{d}y\,y,
\end{align}

Equipped with these expressions we can convert all quantities appearing in the previous section into Euclidean quantities and, once numerically evaluated, directly compare them with the one obtained in the lattice study of~\cite{Skullerud:2003qu}.

\subsection{Numerical results}
In this subsection we carry out the numerical evaluation of the various relevant quantities introduced so far,
and we compare our results with the lattice data on the quark-gluon vertex. 

\subsubsection{Ingredients}

For the evaluation of the one-loop dressed scalar functions $\K{i}{}$ we need 
the following ingredients : \n{i} the gluon propagator $\Delta$, 
\n{ii} the ghost dressing function $F$, 
\n{iii} the value of the strong coupling, at the relevant renormalization scale, $\mu$. Specifically, since 
the lattice data on the quark-gluon vertex have been renormalized at $\mu=2.0$ GeV~\cite{Skullerud:2003qu}, 
this particular scale will serve as our reference,
and all quantities will be renormalized, for consistency, at this particular point.
\n{iv} 
the Dirac vector and scalar components of the quark propagator, $A$ and $B$, respectively.
In what follows we 
explain briefly how the above ingredients are obtained.

\begin{itemize}
\item[\n{i}] 
As in a variety of previous 
works~(e.g.,~\cite{Aguilar:2010cn,Aguilar:2011ux,Aguilar:2010gm,Aguilar:2011yb}), we use  
for the gluon propagator $\Delta$ directly
the SU(3) lattice data of~\cite{Bogolubsky:2009dc}.
As has been explained in detail in the 
literature cited above, an excellent, physically motivated fit of the lattice data 
(renormalized at $\mu=4.3$ GeV, the last available point in the ultraviolet tail of the gluon propagator), 
is given by  
\be
\Delta^{-1}(q^2)= M^2(q^2) + q^2\left[1+ \frac{13C_{\rm A}g_1^2}{96\pi^2} 
\ln\left(\frac{q^2 +\rho_1\,M^2(q^2)}{\mu^2}\right)\right],
\label{gluon}
\ee  
where
\be
M^2(q^2) = \frac{m_0^4}{q^2 + \rho_2 m_0^2}.
\label{dmass}
\ee
Notice that in the above expression, the finiteness of $\Delta^{-1}(q^2)$ is assured  
by the presence of the function $M^2(q^2)$, which forces the value 
of \mbox{$\Delta^{-1}(0) = M^2(0) = m_0^2/\rho_2$}.
The best fit obtained with this functional form corresponds to 
setting \mbox{$m_0 = 520$~MeV}, $g_1^2=5.68$, $\rho_1=8.55$  and  $\rho_2=1.91$. 

Of course, since we want our results renormalized at $\mu=2.0$ GeV instead of $\mu=4.3$ GeV,  
the curve of \1eq{gluon} must be rescaled by a multiplicative factor.  
This factor can be obtained from the standard relation
\be
\Delta(q^2,\mu^2)=\frac{\Delta(q^2,\nu^2)}{\mu^2\Delta(\mu^2,\nu^2)} \,, 
\label{ren_gl}
\ee
which allows one to convert a set of points renormalized at $\nu$ 
to the corresponding set renormalized at $\mu$. In our case $\nu=4.3$ GeV and $\mu=2.0$ GeV,
and  $\Delta(\mu^2,\nu^2) \approx 0.384$ ${\rm GeV^{-2}}$, so that 
the multiplicative factor is $[\mu^2\Delta(\mu^2,\nu^2)]^{-1}\approx 0.652$ .  
The corresponding fit is shown in~\fig{fig:Delta}. 

%%%%%%%%%%%%%%%%%%
\begin{figure}[!t]
%\mbox{}\hspace{-.4cm}
\centerline{\includegraphics[scale=.64]{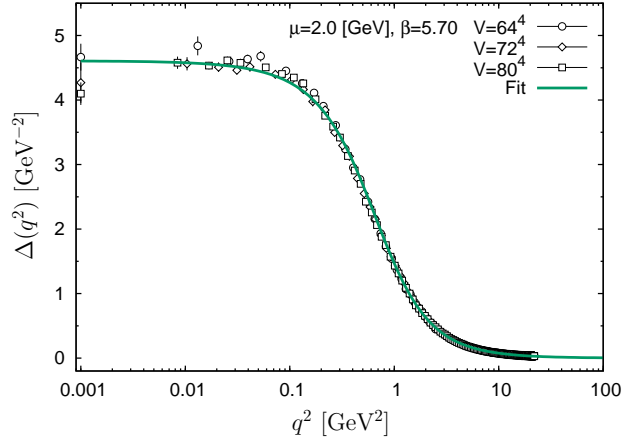}}
\caption{\label{fig:Delta}(color online). 
The functional fit given in~\1eq{gluon} to the SU(3) gluon propagator. Lattice data are taken from~\cite{Bogolubsky:2009dc} and renormalized at $\mu=2.0$~GeV.}
\end{figure} 
%%%%%%%%%%%%%%%%% 

\item[\n{ii},\n{iii}] The ghost dressing function $F$ is determined by solving the 
corresponding ghost gap equation. 
For the fully dressed ghost-gluon vertex entering in it 
we use the expressions obtained in~\cite{Aguilar:2013xqa}. Then, 
the strong coupling $\alpha(\mu^2) = g^2(\mu^2)/4\pi$ is simultaneously fixed by demanding that the solution obtained for $F$ 
matches the SU(3) lattice results of~\cite{Bogolubsky:2009dc}. The best match is achieved for 
 $\alpha=0.45$ at $\mu=2.0$ GeV, as shown in the left panel of \fig{fig:F-1pGandL}. 
From now on $\alpha$ will be kept fixed at this particular value. 
The (inverse) ghost dressing function can be further separated in its $1+G$ and $L$ components that appears in~\1eq{funrel}. This is done by solving the SDEs they satisfy~\cite{Aguilar:2009pp}, and the corresponding results are shown in the right panel of the same figure.

%%%%%%%%%%%%%%%%%%
\begin{figure}[!t]
\centerline{\includegraphics[scale=.995]{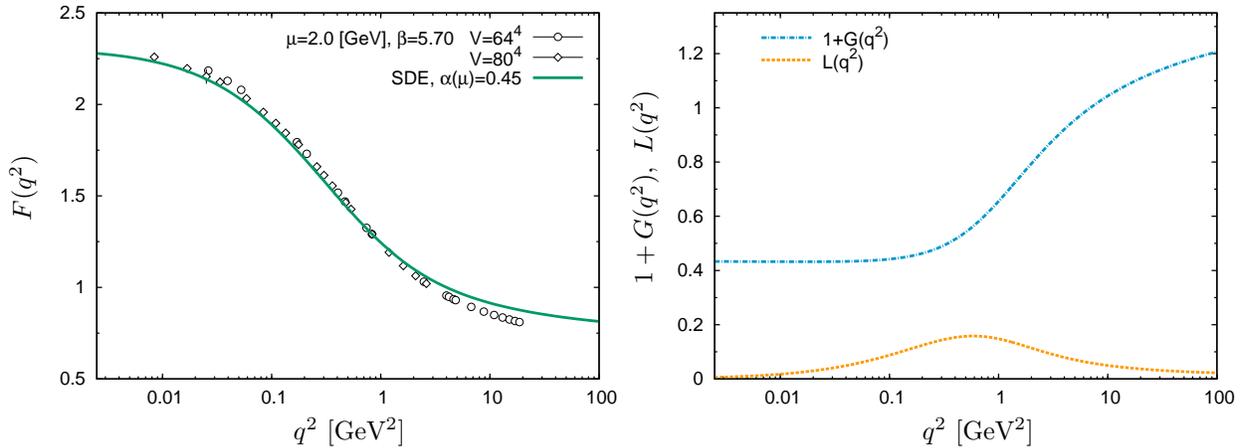}}
\caption{\label{fig:F-1pGandL}(color online). {\it Left}: The Landau gauge ghost dressing function $F$ obtained as a solution of the ghost gap equation for $\alpha=0.45$ using as input the lattice gluon propagator. {\it Right}: The decomposition of the (inverse) ghost dressing function into its $1+G$ (blue, dashed-dotted) and $L$ (orange, dashed) components. Lattice data are taken from~\cite{Bogolubsky:2009dc}.} 
\end{figure} 
%%%%%%%%%%%%%%%%% 

\item[\n{iv}]
With the $\Delta$ and $F$ we have just determined, one can evaluate the vector and scalar components of the quark propagator. This is achieved by 
solving the quark gap equation described in~\cite{Aguilar:2010cn} with a Curtis-Pennington  quark-gluon vertex~\cite{Curtis:1990zs}, 
and a  bare quark mass fixed at 115 MeV, which is the value employed in the lattice simulations of~\cite{Skullerud:2003qu}.  
The results obtained for the quark wavefunction $Z=1/A$ and mass $M=B/A$ are shown in~\fig{fig:AandB_SDE}. 

%%%%%%%%%%%%%%%%%%
\begin{figure}[!t]
\mbox{}\hspace{-.4cm}\includegraphics[scale=.995]{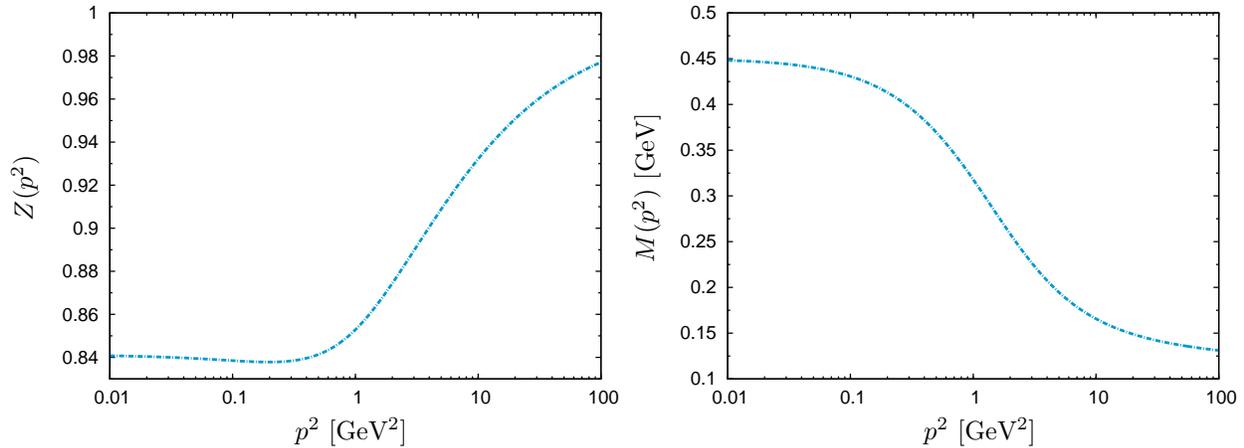}
\caption{\label{fig:AandB_SDE}(color online). 
The quark wave-function (left), and mass (right), obtained from the solution of the quark gap equation for a  current mass  $m_0=115 $ MeV and $\alpha_s(\mu)=0.45$.} 
\end{figure} 
\end{itemize}
At this point we have all the ingredients and shall proceed to determine the $\K{i}{}$ and $\oK{i}{}$ auxiliary functions, and subsequently the vertex form factors for the various kinematical limits introduced before.  

\subsubsection{{\label{soft_mum}}Soft-gluon limit}

\begin{figure}[!b]
\centerline{\includegraphics[scale=.64]{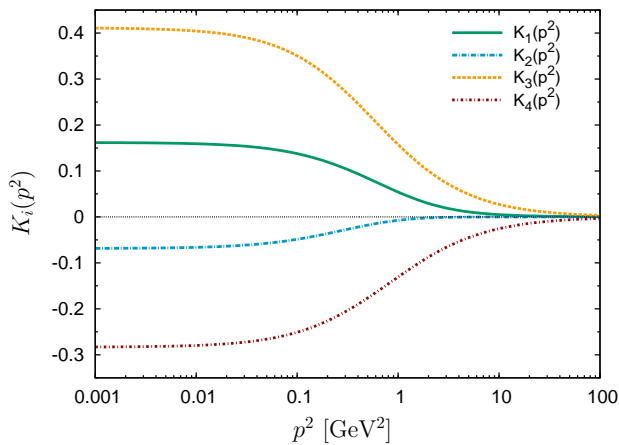}}
\caption{\label{fig:theKs_soft}(color online). The auxiliary functions $K_i$ evaluated in the soft gluon limit.} 
\end{figure}

In \fig{fig:theKs_soft} we plot the functions $K_i$ in the soft gluon limit~\1eq{theKs}, obtained using $\Delta$, $F$, $Z$ and $M$ determined in the previous section. 

%%%%%%%%%%%%%%%%%%
%%%%%%%%%%%%%%%%% 

\begin{figure}[!t]
\centerline{\includegraphics[scale=.995]{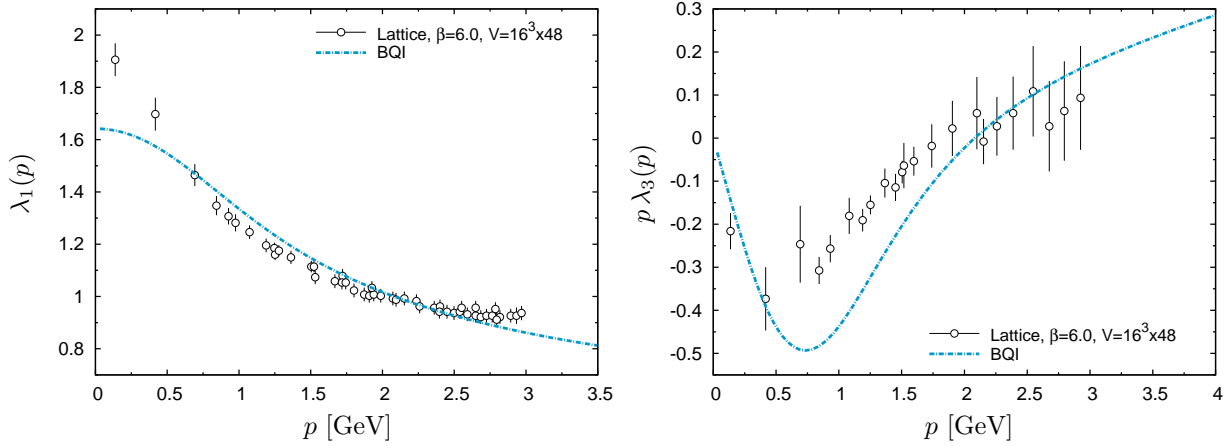}}
\caption{\label{fig:Gamma1-Gamma3_soft}(color online). 
The soft gluon form factors $\lambda_1$ (left) and $p\lambda_3$ (right). Lattice data in this and all the following plots are taken from~\cite{Skullerud:2003qu}.} 
\end{figure}

It is then immediate to construct the Euclidean version of the soft gluon limit form factors~\noeq{softlimit-results-Mink}. 
Specifically, in~\fig{fig:Gamma1-Gamma3_soft} we plot the form factors
\begin{equation}
\lambda_1(p)=\Ga{1}{\s{\rm E}}(p_\s{\rm E});\qquad 
\lambda_3(p)=-\frac12\Ga{3}{\s{\rm E}}(p_\s{\rm E}),
\end{equation}
and compare them with the lattice data of~\cite{Skullerud:2003qu}, obtaining a rather satisfactory agreement.  

%%%%%%%%%%%%%%%%%%
%%%%%%%%%%%%%%%%% 

\begin{figure}[!b]
\centerline{\includegraphics[scale=.64]{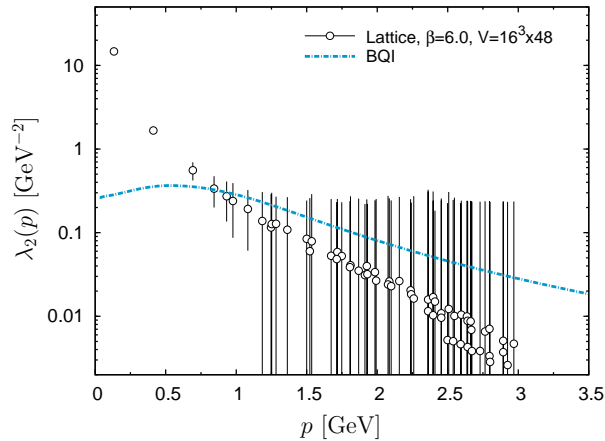}}
\caption{\label{fig:Gamma2_soft}(color online). The form factor $\lambda_2$ and the corresponding lattice data.} 
\end{figure}

However, in the case of the form factor
\begin{equation}
\lambda_2(p)=\frac14\Ga{2}{\s{\rm E}}(p_\s{\rm E}),
\end{equation}
we observe a fundamental qualitative discrepancy with respect to the lattice data; in particular, as \fig{fig:Gamma2_soft} shows, we obtain a finite form factor, while the lattice shows an IR divergence as $p^2\to0$. %To be sure the data are well described by $\lambda_2/p^2$ (see the gray curve in~\fig{fig:Gamma2_soft}). 

%%%%%%%%%%%%%%%%%%
%%%%%%%%%%%%%%%%% 

This discrepancy seems common to all attempts to evaluate the quark-gluon vertex form factors from a 
purely SDE approach (see for example~\cite{Bhagwat:2004kj,LlanesEstrada:2004jz}). In what follows we  
will offer a plausible explanation for its origin, at least within our framework.

To begin with, recall that in the soft gluon limit the transverse parts of the vertex are not active; 
thus, the expressions~\noeq{softlimit-results-Mink} are exact, and any divergence can only manifest 
itself in the auxiliary functions $K_i$. Specifically,~\1eq{softlimit-results-Mink} shows that in $\lambda_2$ 
only the functions $K_i$ with $i=2,3,4$ appear. In general, however, $K_3$ should not develop a  
IR divergence, since this would render IR divergent also $\lambda_3$, and we know from the 
lattice that this form factor is finite (see~\fig{fig:Gamma1-Gamma3_soft}). 
On the other hand, both $K_2$ and $K_4$ could in principle have 
an IR divergence, as long as they diverge at most as $1/p^2$, 
given that they both appear in $\lambda_1$ and $\lambda_3$ multiplied by a factor $p^2$.

To analyze what happens in the $p\to0$ limit of these two functions, let us  
observe that in the soft gluon limit the function $R^f$ of~\1eq{Rf-soft-symm} can be written as 
\begin{equation}
R^f(k,p)=\Delta(k^2)D(k^2)g(k+p);\qquad g(k+p)=\frac{f(k+p)}{A^2(k+p)(k+p)^2-B^2(k+p)}.
\end{equation}
The function $g$ can be next expanded around $p=0$ according to
\begin{equation}
g(k+p)=g(k^2)+2(k\cd p)g'(k^2)+p^2g'(k^2)+2(k\cd p)^2g''(k^2)+{\cal O}(p^3),
\label{exp}
\end{equation}
where the primes denote derivatives w.r.t. $k^2$. All functions appearing in the above expansion of $g$ are well behaved in the IR, 
and we will assume the same about their derivatives.

One may then establish that the $K_4$ in \1eq{theKs}
is regular as $p\to0$; indeed, the only possible divergence may come from the zeroth order term in~\noeq{exp}.
This term, however, vanishes, since it is proportional to the integral of  
$(k\cd p)g(k^2)/p^2$, which is an odd function of the integration angle $\theta$ in the interval $[0,\pi]$.
In the case of $K_2$, the presence of the prefactor $1/p^2$ implies that one has to consider 
both the zeroth and the first order term in the expansion~\noeq{exp}. 
Again, however, they both vanish: the linear term in $p$ for the same reason as before (odd in $\theta$) , 
while the zeroth order term due to the vanishing of the corresponding angular integral, namely\footnote{Note that if the expressions 
in \1eq{theKs} are worked out 
in $d$ space-time dimensions, one obtains the factor ($1-d\cos^2\theta)$; thus, 
the result of \1eq{angaccsymm} is particular to $d=4$.}
\begin{equation}
\int_0^\pi\!\mathrm{d}\theta\,\sin^2\theta(1-4\cos^2\theta)\, =0. 
\label{angaccsymm}
\end{equation} 
As a result, the one-loop dressed $K_2$ and $K_4$ in the soft gluon limit both saturate to a constant in the IR, as~\fig{fig:theKs_soft} shows.

%%%%%%%%%%%%%%%%%%
\begin{figure}[!t]
\centerline{\includegraphics[scale=0.995]{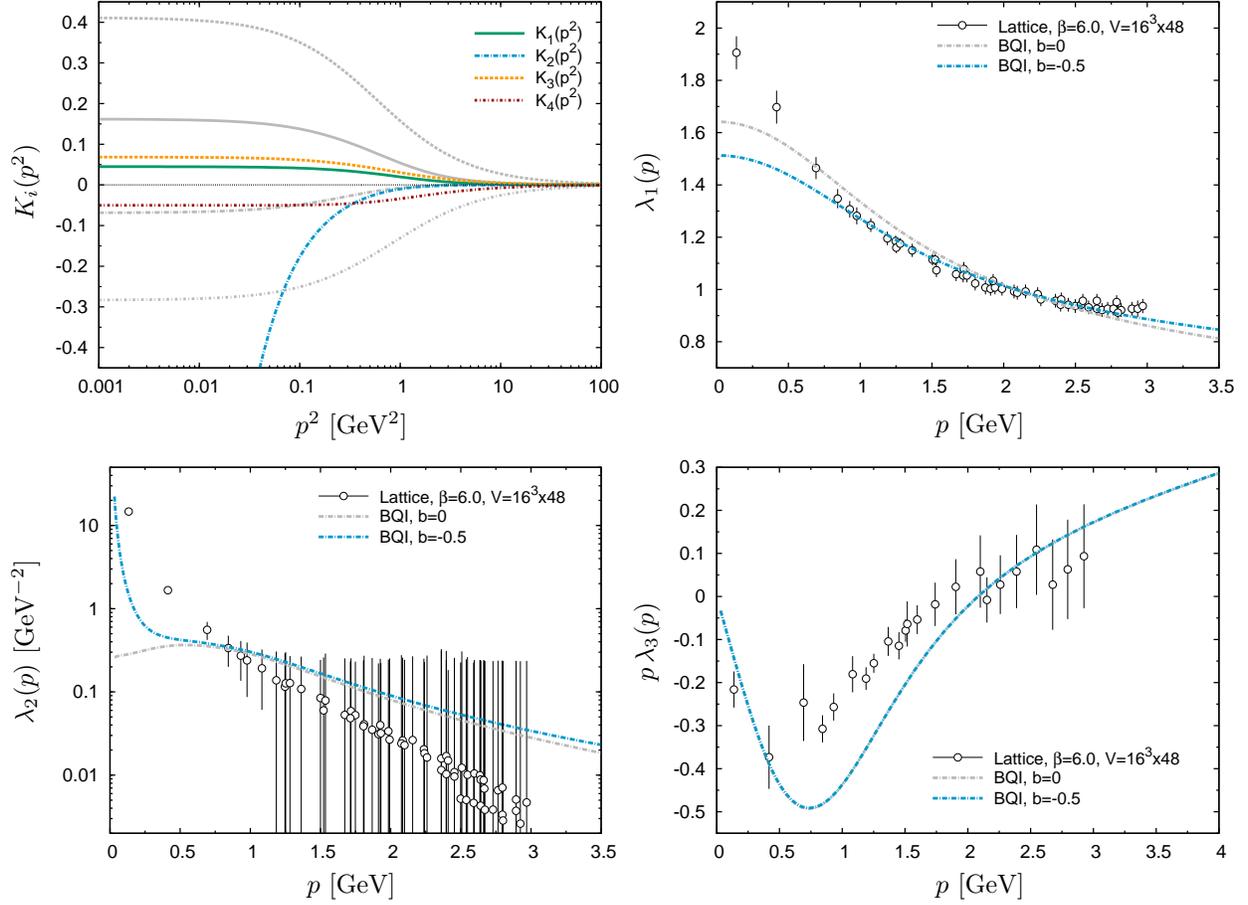}}
\caption{\label{fig:cos2theta}(color online). The auxiliary functions $K_i$ evaluated in the soft gluon limit with a fermion vertex $\cos^2\theta\gamma_\mu$. When comparing with the results obtained for the tree-level vertex $\gamma_\mu$ (gray curves) one notice that $K_2$ becomes IR divergent, whereas the remaining $K_i$ are suppressed. In the remaining panels we show the soft gluon form factors obtained when using the vertex \mbox{$\gamma_\mu(1+b\cos^2\theta)$} for the representative value $b=-0.5$; one obtains a divergent $\lambda_2$, affecting only modestly $\lambda_1$ and leaving $\lambda_3$ practically invariant.} 
\end{figure} 
%%%%%%%%%%%%%%%%% 

Evidently, the finiteness of the form factor $K_2$ in the soft gluon limit originates from the conspiracy 
of two independent facts: 
\n{i} The IR finiteness of the expanded function $g$, 
which implies that the ${\cal O}(p^2)$ terms in~\1eq{exp} will give rise to an IR convergent integral.
Instead, in the symmetric and zero quark momentum limits, 
the function to be expanded involves always the IR divergent ghost propagator, and therefore $K_2$ will be IR divergent in both cases
(see Figs~\ref{fig:theKs_symm} and~\ref{fig:theKs-thebarKs_zero}).
\n{ii}  The vanishing of the angular integral~\noeq{angaccsymm} (in 4 space-time dimensions). 

Now, the presence of the integral~\noeq{angaccsymm} can be traced back to the 
one-loop dressed approximation we have used to evaluate the functions $K_i$, where the 
fermion vertex was kept at tree-level. In that sense, the obtained finiteness of $K_2$ is accidental, being
really an artefact of our particular implementation of the one-loop dressed approximation.  
Actually, if one were to include some additional angular dependence 
to this vertex (which will happen anyway when quantum corrections are added), 
the cancellation~\noeq{angaccsymm} would be unavoidably distorted, and one would end up with an IR divergent $K_2\sim1/p^2$.

This fact is shown in~\fig{fig:cos2theta}, where the tree-level vertex $\gamma_\mu$ has been replaced by \mbox{$\gamma_\mu(1+b\cos^2\theta)$}. 
One observes that $K_2$ becomes indeed IR divergent as soon as $b\neq0$, 
while all remaining $K_i$ are only modestly affected by the presence of $b$. 
The resulting form factors for the representative value $b=-1/2$ 
are shown in the same figure: $\lambda_2$ develops a $1/p^2$ IR divergence, while $\lambda_1$ and $\lambda_3$ are marginally modified.

\subsubsection{Symmetric limit}

%%%%%%%%%%%%%%%%%%
\begin{figure}[!t]
\centerline{\includegraphics[scale=.65]{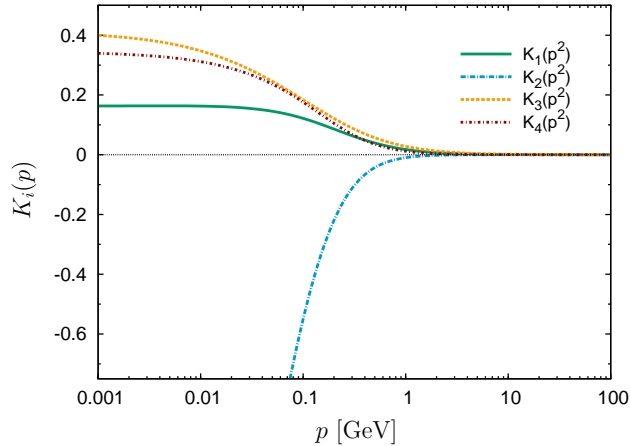}}
\caption{\label{fig:theKs_symm}(color online). The auxiliary functions $K_i$ evaluated in the symmetric gluon limit when an extra angular dependence of the type $b\cos^2\theta$ is added to the tree-level vertex $\gamma_\nu$.} 
\end{figure} 
%%%%%%%%%%%%%%%%% 

Let us now turn our attention to the symmetric limit. According to our previous discussion, in this limit
we expect a divergent $K_2$ and a finite $K_4$, 
as indeed shown in~\fig{fig:theKs_symm}.

%%%%%%%%%%%%%%%%%%
\begin{figure}[!t]
\centerline{\includegraphics[scale=.995]{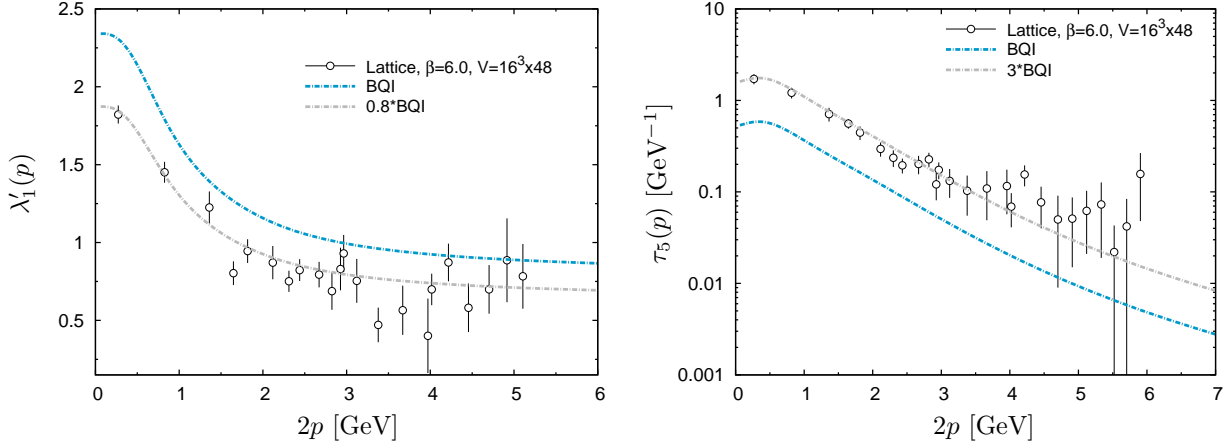}}
\caption{\label{fig:lambda1-tau5_symm}(color online). 
The symmetric limit form factors $\lambda'_1$ and $\tau_5$ compared with the corresponding lattice data.
The grey curves are obtained through simple rescaling of the blue ones.} 
\end{figure} 
%%%%%%%%%%%%%%%%% 

We next proceed to plot (\fig{fig:lambda1-tau5_symm}) the form factors 
\begin{equation}
\lambda'_1(p)=\Ga{1}{L\s{\rm{E}}}(p_\s{\rm{E}})-p^2_\s{\rm{E}}\Ga{3}{T\s{\rm{E}}}(p_\s{\rm{E}});\qquad 
\tau_5(p)=\frac12\Ga{5}{T\s{\rm{E}}}(p_\s{\rm{E}}),
\label{prj-symm}
\end{equation}
which are measured on the lattice in the symmetric limit. As~\1eq{symmff} shows, these form factors do not explicitly involve the divergent term $K_2$, and therefore are finite. For both of them the overall shape of the lattice data is 
accurately described; however, the strength of the two components is inverted, 
since we get a higher $\lambda'_1$ and a lower $\tau_5$. 
Quite interestingly, a simple rescaling of each form factor (through multiplication by a numerical constant) 
leads to a very good overlap with the lattice data, as the gray curves demonstrate.

%%%%%%%%%%%%%%%%%%
\begin{figure}[!b]
\centerline{\includegraphics[scale=.995]{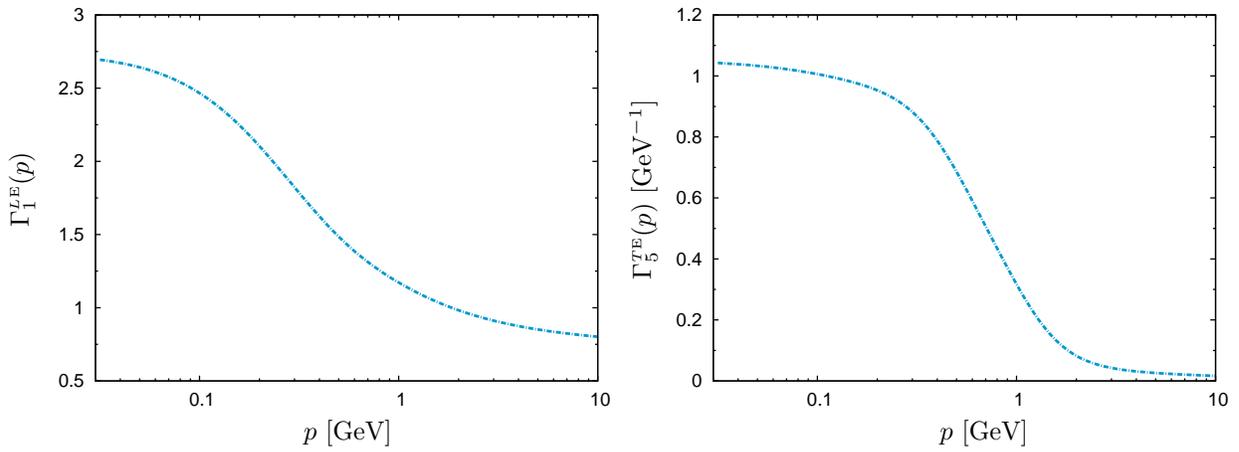}}
\caption{\label{fig:Gamma1L-Gamma5T_symm}(color online). 
The symmetric form factors $\Ga{1}{L}$ (left) and $\Ga{5}{T}$ (right).} 
\end{figure} 
%%%%%%%%%%%%%%%%% 

%%%%%%%%%%%%%%%%%%
\begin{figure}[!t]
\centerline{\includegraphics[scale=.65]{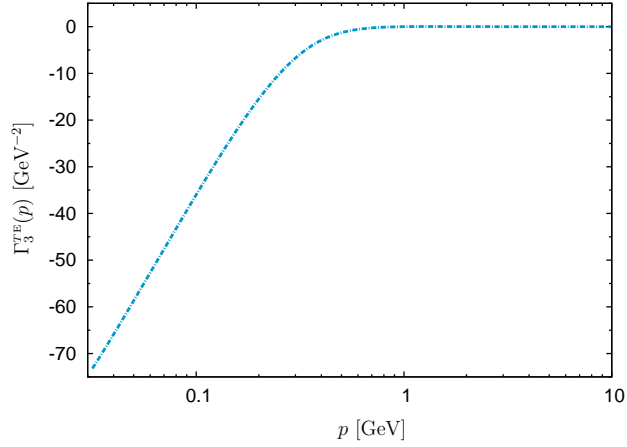}}
\caption{\label{fig:Gamma3T_symm}(color online). The divergent form factor $\Ga{3}{T}$ in the symmetric limit.} 
\end{figure} 
%%%%%%%%%%%%%%%%% 

Our analysis is not limited to the projected form factors~\noeq{prj-symm}, 
as we can study also all the three non-zero form factors~\noeq{symlimit-results-M} 
in this limit, similarly to what we have done in the soft gluon limit. 
Specifically, on the basis of~\1eq{symlimit-results-M} one expects that $\Ga{1}{\s L}$ and $\Ga{5}{\s T}$ 
are finite (\fig{fig:Gamma1L-Gamma5T_symm}), as they involve only the combination $p^2K_2$ through the term $\K{1}{L}$; 
however, $\Ga{3}{\s T}$ has two divergent pieces~(\fig{fig:Gamma3T_symm}), 
both proportional to the combination $L_{2p}/p^2$ 
reading\footnote{In the SDE for the function $L$, the ghost-gluon vertex has been approximated by its tree-level value; 
however, the dressing of this vertex is not expected to alter the above argument}~\cite{Aguilar:2009pp}
\begin{equation}
\frac1{p^2}L_{2p}\sim\frac1{p^2}\int_k \left[1 - d \, \frac{(k \cdot p)^2}{k^2 p^2}\right]\Delta (k)  D(k+2p)\underset{p^2\to0} {\sim}\frac1p.
\end{equation}

\subsubsection{Zero quark momentum}

The case of the zero quark momentum constitutes a ``prediction'', given that there are no lattice data available for this particular momentum configuration.

%%%%%%%%%%%%%%%%%%
\begin{figure}[!b]
\centerline{\includegraphics[scale=.995]{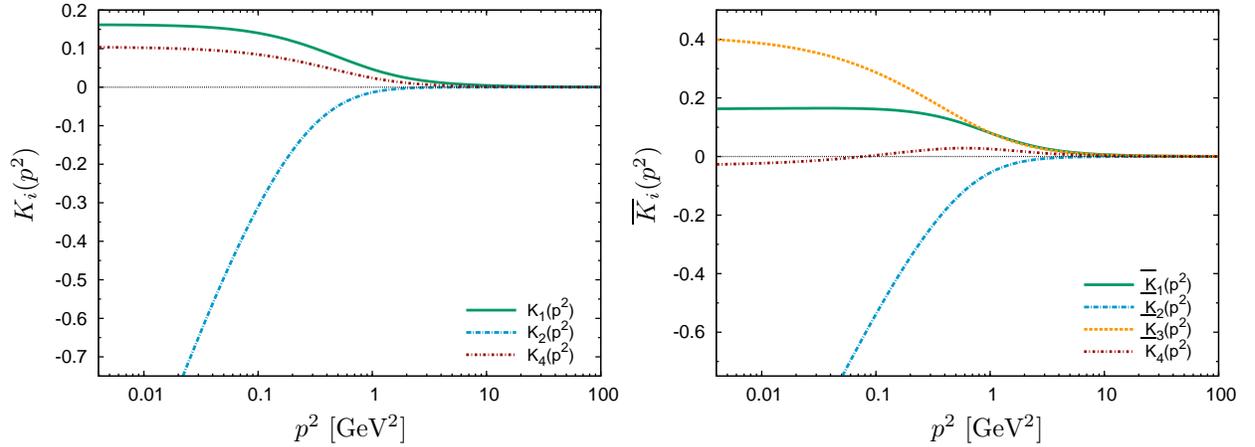}}
\caption{\label{fig:theKs-thebarKs_zero}(color online). 
The auxiliary functions $\K{i}{}$ (left) and $\oK{i}{}$ (right) evaluated in the zero quark momentum configuration.} 
\end{figure} 
%%%%%%%%%%%%%%%%% 

In this specific case, the degeneracy between the $\K{i}{}$ and $\oK{i}{}$ functions is broken, and one has to study them separately. In addition, one will have both $\K{2}{}$ and $\oK{2}{}$ divergent in this case, even though the (projected) form factors $\lambda'_1$ and $\tau_5$, introduced in~\1eq{prj-symm}, 
will still be finite, as the only combination that enters in their definition~\noeq{ffzero} is $\K{1}{L}=\K{1}{}+p^2\K{2}{}$.

%%%%%%%%%%%%%%%%%%
\begin{figure}[!t]
\centerline{\includegraphics[scale=.995]{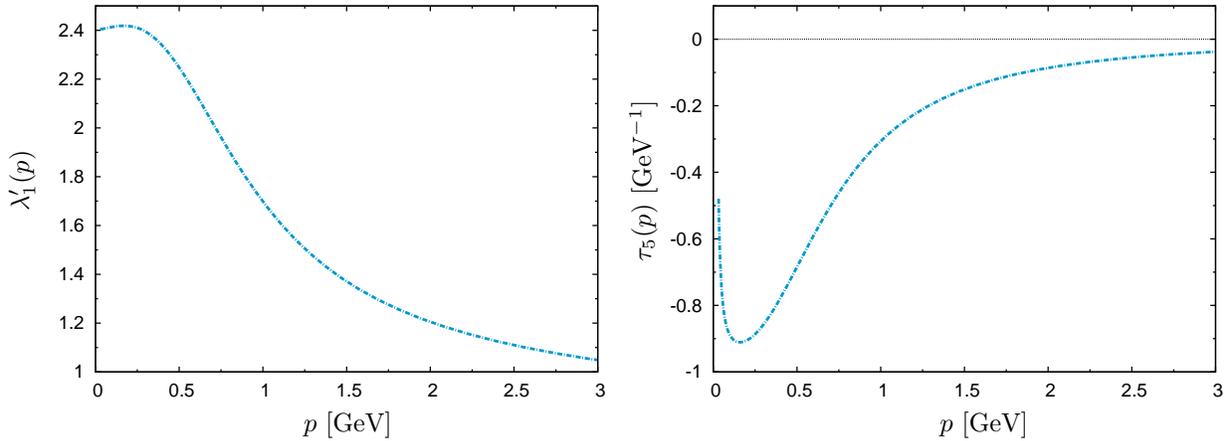}}
\caption{\label{fig:lambda1-tau5_zero}(color online). 
The form factors $\lambda'_1$ (left) and $\tau_5$ (right) in the quark zero momentum configuration.} 
\end{figure} 
%%%%%%%%%%%%%%%%% 

\begin{figure}[!t]
\mbox{}\hspace{-1.3cm}\includegraphics[scale=.995]{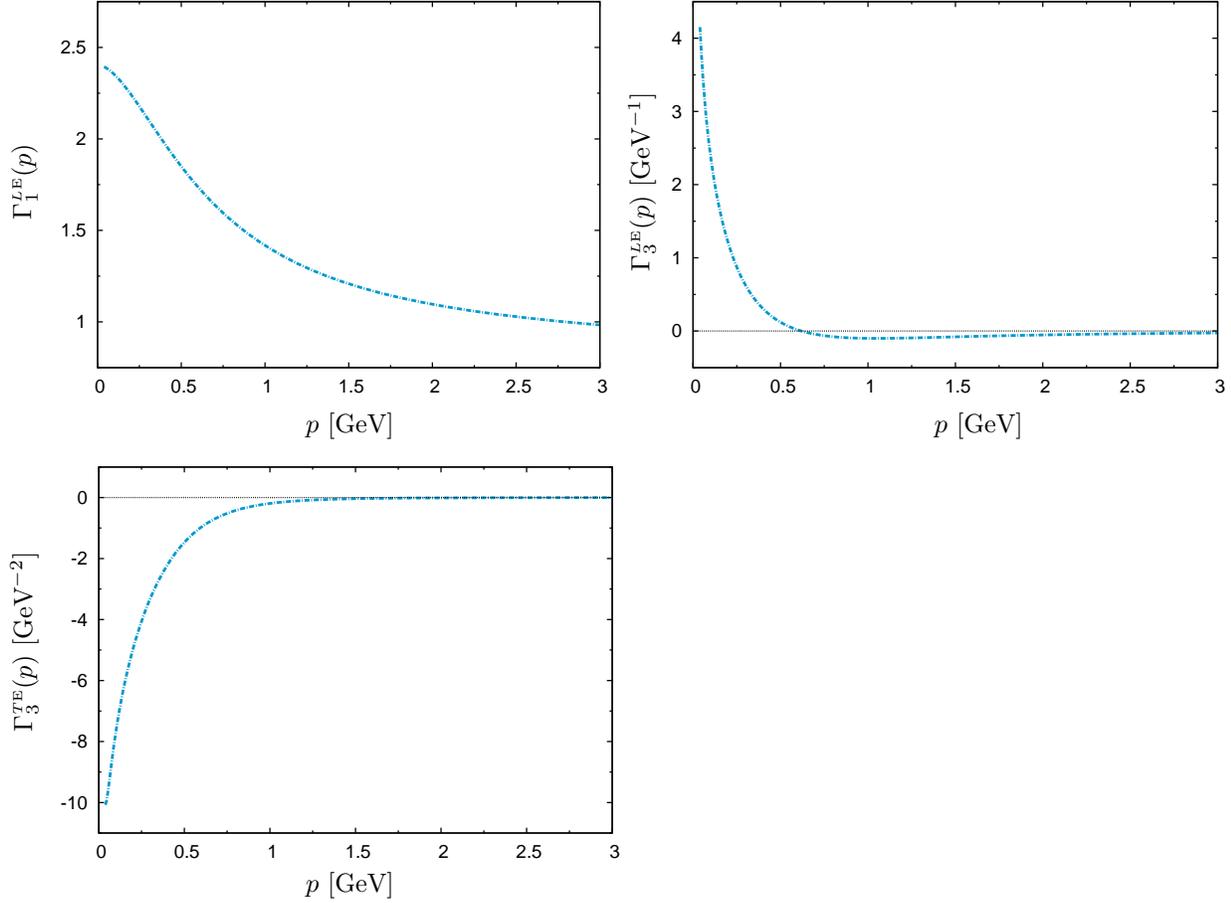}
\caption{\label{fig:allGamma_zero}(color online). 
The form factors $\Ga{1}{L\s{\rm E}}$, $\Ga{3}{L\s{\rm E}}$ and $\Ga{3}{T\s{\rm E}}$ evaluated in the zero quark momentum configuration.} 
\end{figure}

In~\fig{fig:theKs-thebarKs_zero} we plot the auxiliary functions $\K{i}{}$ and $\oK{i}{}$, 
while in~\fig{fig:lambda1-tau5_zero} we plot the form factors defined in~\1eq{prj-symm}, 
which, in principle, could be simulated on the lattice. Finally, in \fig{fig:allGamma_zero}, 
we present all form factors; notice in particular the (negative) divergence expected for the term $\Ga{3}{T\s{\rm E}}$.

%%%%%%%%%%%%%%%%%%
%%%%%%%%%%%%%%%%% 

\subsection{Unquenching effects}

An additional issue worth mentioning is related with the fact that the
lattice results  that we  have been  using as  initial ingredients
(gluon  propagator  and ghost  dressing  function)  are obtained  from
quenched  simulations  (no dynamical  quarks).  To be sure, 
the  procedure of using quenched  results to
obtain  dynamical properties  of quarks  may be considered,  strictly speaking,
inconsistent. However, from the practical point of view, 
it has been argued in earlier works~\cite{Bhagwat:2004kj,LlanesEstrada:2004jz}
that the effects of unquenching are relatively small (of the order of $10\%$), and may be omitted 
as a first approximation.

 To show that the above error estimate  is valid also within our approach, we  repeat again the  
soft gluon configuration  analysis, but now  using the unquenched  gluon and ghost  
lattice propagators obtained in~\cite{Ayala:2012pb}, and the correspondingly modified values 
for the strong coupling constant $\alpha$.
 More specifically, we compute the form factors $\lambda_1(p)$, $\lambda_2(p)$ and $\lambda_3(p)$ 
for two different numbers of active flavors (i) $N_f=2$
(two degenerate light quarks) and (ii) $N_f=2+1+1$ (two degenerate
light quarks and two heavy ones), which were considered in the lattice simulations of~\cite{Ayala:2012pb}.
The corresponding values of $\alpha(\mu)$ (at $\mu=2$\, GeV) are obtained by repeating the same procedure outlined in the first 
subsection of this section (items $\n{ii},\n{iii}$); specifically we have \mbox{$\alpha(\mu)=0.45$} for \mbox{$N_f =0$}, 
\mbox{$\alpha(\mu)=0.59$} for \mbox{$N_f =2$}, and  \mbox{$\alpha(\mu)=0.66$} for\mbox{$N_f =2+1+1$}.

\begin{figure}[!t]
\centerline{\includegraphics[scale=.995]{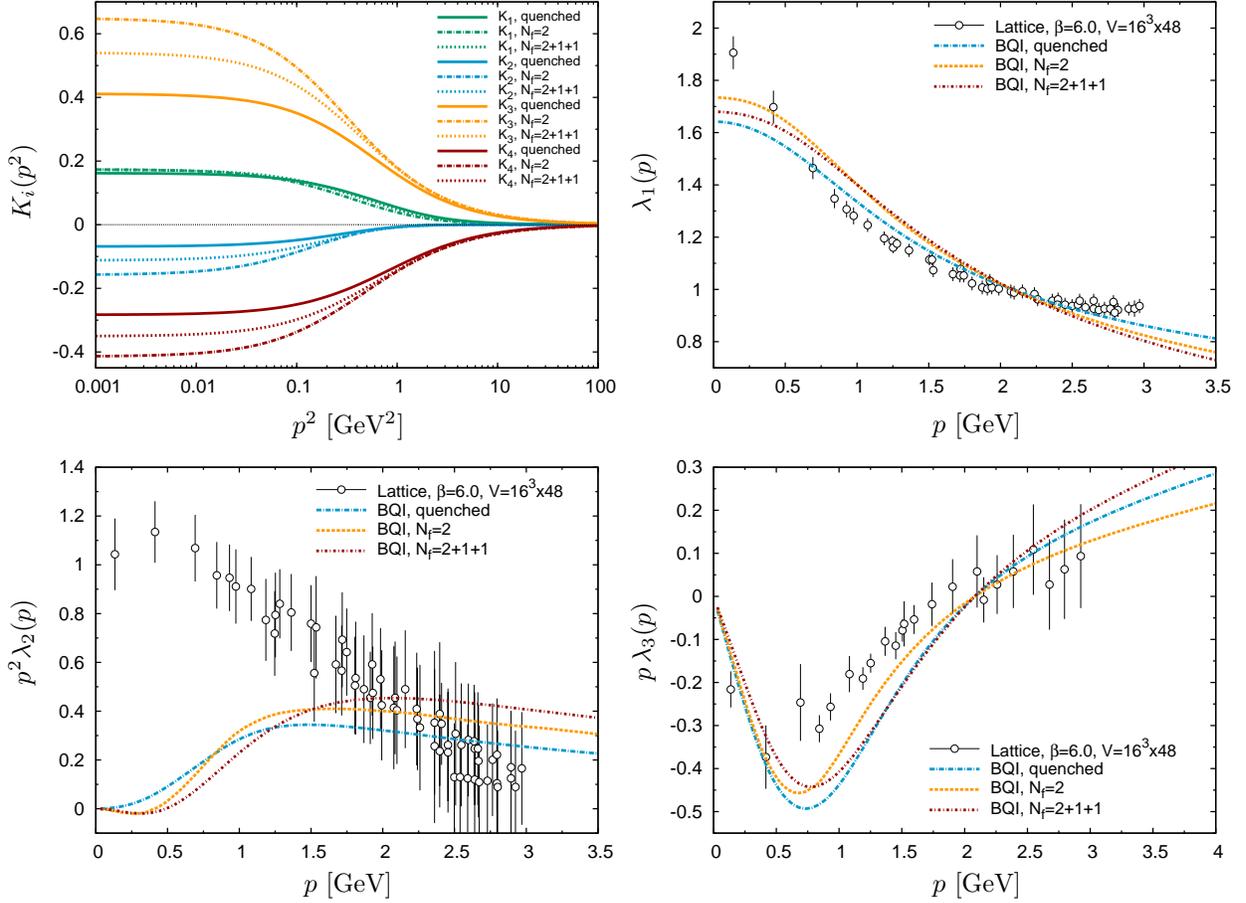}}
\caption{\label{fig:soft_comp1}(color online). 
The comparison of the results for $N_f=0$, $N_f=2$ and $N_f=2+1+1$ in the
soft gluon configuration. We show  the  auxiliary functions $K_i$ (upper left panel) and the form factors: $\lambda_1(p)$ (upper right), $p^2\lambda_2(p)$ (bottom left) and $p\lambda_3(p)$ (bottom right).} 
\end{figure}

Our results for the soft gluon configuration, using the unquenched propagators as input,  are shown in Fig.~\ref{fig:soft_comp1}. 
On the upper left panel 
we plot the functions $K_i$, whereas on the right one we show the $\lambda_1(p)$ 
for the quenched case (blue), $N_f=2$ (orange), and $N_f=2+1+1$ (red). Using the same color code, we show,  
on the bottom panels $p^2\lambda_2(p)$ (left) and $p\lambda_3(p)$ (right), respectively.
Evidently, the effect of activating the quarks 
amounts to scaling up the quenched result by less than $6\%$ (in the deep IR region).
The same type of quantitative changes are observed for  $p^2\lambda_2(p)$. 
Finally, in the case of $p\lambda_3(p)$, the biggest difference between quenched and unquenched cases  occurs when \mbox {$ p\sim 1$ GeV} and is about $10\%$.

\subsection{Comparison with previous works}

%%%%%%%%%%%%%%%%%
%%%%%%%%%%%%%%%%% 

%%%%%%%%%%%%%%%%%
\begin{figure}[!t]
\centerline{\includegraphics[scale=.995]{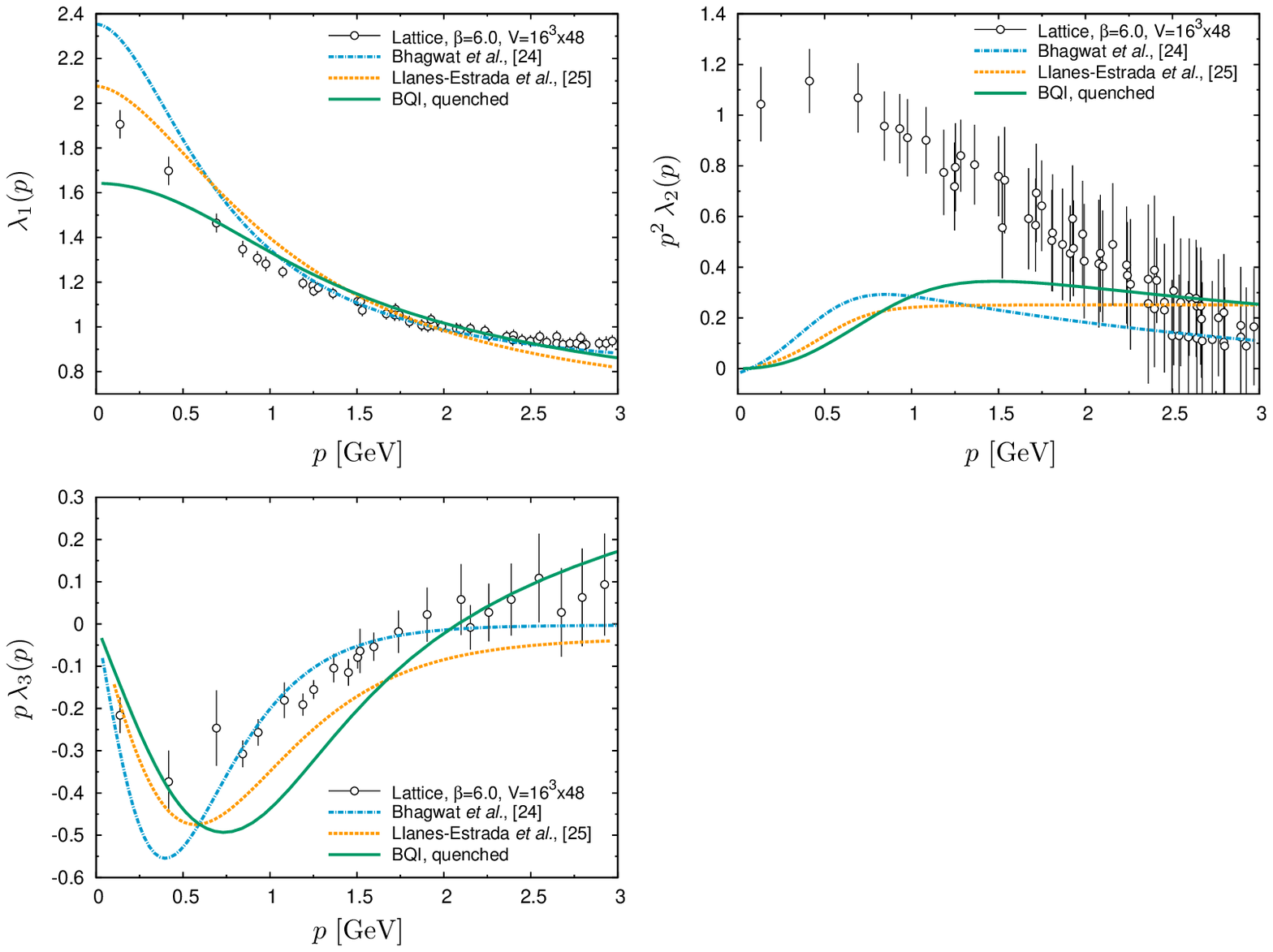}}
\caption{\label{fig:soft_comp2}(color online). 
Our quenched results for the  soft gluon configuration (green, continuous) compared with the results obtained in Refs.~\cite{Bhagwat:2004kj} (blue, dashed-dotted) and~\cite{LlanesEstrada:2004jz} (orange, dashed).} 
\end{figure} 
%%%%%%%%%%%%%%%%% 

For completeness, in this subsection we compare our results with those 
obtained by two different approaches, representative of the extensive literature on this 
subject.
Specifically, 
in order to carry out a concrete comparison, we will concentrate on 
the contributions presented in Refs.~\cite{Bhagwat:2004kj,LlanesEstrada:2004jz}. In particular, in Fig.~\ref{fig:soft_comp2} we    
compare our results for the soft gluon form factors $\lambda_1(p)$, $p^2\lambda_2(p)$, and $p\lambda_3(p)$  with those obtained using
two different semiperturbative analysis presented in  Refs.~\cite{Bhagwat:2004kj} and~\cite{LlanesEstrada:2004jz}.

In both aforementioned works the form factors of the  quark-gluon vertex are obtained  
in the context of the one-loop dressed approximation. 
In the case of~\cite{Bhagwat:2004kj}, the two relevant diagrams (in the  soft gluon  limit)
were calculated within the ``rainbow-ladder'' approximation, 
using for the product   $g^2\Delta(q^2)$ a phenomenological model frequently employed    
in Bethe-Salpeter studies~\cite{Maris:1999nt}. 
In addition, all bare quark propagators 
are replaced by the solutions of the quark SDE, obtained in the same rainbow approximation. 
On the other hand, in~\cite{LlanesEstrada:2004jz} the relevant diagrams
were computed by replacing the internal tree-level quark and 
gluon propagators by their dressed counterparts,  
calculated in the ghost dominance picture~\cite{Fischer:2003rp}.
There, the authors analysed the vertex in 
the {\it (i)} soft gluon limit and in the {\it (ii)} totally asymmetric configuration. 

It is interesting to observe that the three studies compared here display the  
same qualitative behavior for $p^2\lambda_2(p)$ at one-loop dressed approximation: $p^2\lambda_2(p)$
tends to zero in the deep IR region, while the lattice data is clearly finite. T
he above observation reinforces the arguments presented 
in the Sec.~\ref{soft_mum}, where the finiteness of the kernel $\lambda_2(p)$ was interpreted as an ``artefact'' 
of a special numerical cancellation operating  
at the level of one-loop dressed approximation.

\section{\label{sec:concl}Discussion and Conclusions}

In this article we have presented a novel method for determining the 
nonperturbative quark-gluon vertex, which constitutes a crucial ingredient  
for a variety of theoretical and phenomenological studies. 
Our method is particular to the PT-BFM scheme, and relies heavily 
on the BQI relating $\Gamma_{\mu}$ and $\widehat\Gamma_{\mu}$. The  
TWIs are of paramount importance in this approach, because they provide nontrivial information 
on the transverse part of $\widehat\Gamma_{\mu}$ (and eventually of $\Gamma_{\mu}$).    

One important difference of this method compared to the standard SDE approach 
is that it takes full advantage of the rich 
amount of information originating from the fundamental underlying symmetries,
before actually computing (fully-dressed) Feynman diagrams.
In particular, both the BRST and antiBRST symmetries are properly 
exploited, by appealing to a set of crucial identities (WIs, STIs, BQIs), 
in order to obtain nontrivial information for all form-factors, already at the 
first level of approximation. The actual calculation of diagrams is then 
reduced to the auxiliary three-point functions, which have a 
simpler structure compared to the standard SDE expansion. Note in particular that,  
at the level of approximation that we work, the three-gluon vertex, 
a known source of technical complexity, does not enter at all. 
On the other hand, a major downside of this method 
is that the minimal Ansatz employed at the level of the TWI may be hard to improve upon, 
given the nonlocal nature of the omitted terms.

It is important to emphasize at this point that, even though the minimal Ansatz used for the vertex  
satisfies the correct constraints imposed by the general symmetries of the theory, 
its quantitative contribution to the final answer is not necessarily dominant. 
In fact, our analysis reveals that the contributions originating from the one-loop dressed diagrams 
used to calculate $K_{\mu}$ are in general sizeable, and tend to drive the answer towards 
the direction of the lattice results. Therefore, the calculation of these terms, within an 
approximation scheme as refined as possible, is of paramount importance 
for the successful implementation of this particular approach.

The main external ingredient used in the calculation  
of the three-point function $K_{\mu}$
is the nonperturbative gluon propagator $\Delta(q^2)$, 
which has been taken from the lattice.
On the other hand, the ghost dressing function $F(q^2)$ 
and the Dirac components of the quark propagator [$A(p^2)$ and $B(p^2)$] are obtained from the solution of the corresponding SDEs.
To be sure, a completely self-contained analysis ought to include the 
dynamical determination of $\Delta(q^2)$ from its own SDE; 
however, this task is beyond our present powers, mainly due to the poor knowledge of one 
of the ingredients of this SDE, namely the fully dressed four-gluon vertex of the PT-BFM.
 
In general,  
the numerical results presented here appear to be in qualitative agreement  
with those obtained from lattice simulations, following the overall trend 
of the data, but they do not succeed in    
achieving a particularly noteworthy level of quantitative coincidence. 
In the case of the ``soft-gluon limit'', the form-factors $\lambda_{1}(p)$ and 
$p\lambda_{3}(p)$ (shown in Fig.~\ref{fig:Gamma1-Gamma3_soft}) capture clearly the general 
pattern of the lattice results; however, 
$\lambda_{1}(p)$ deviates about 25\% in the deep infrared, while 
$p\lambda_{3}(p)$ shows its largest discrepancy (a factor of about 1.5)
in the region of momenta around 0.75 GeV. The case of $\lambda_{2}(p)$, shown in Fig~\ref{fig:Gamma2_soft}, 
merits particular attention.
Specifically, whereas the one-loop approximation gives a finite answer at the origin
(contrary to the lattice results), a possible mechanism for overcoming this 
has been identified; a divergent result may indeed be obtained (see Fig~\ref{fig:cos2theta}), 
at the expense of introducing an additional parameter ($b$). Note, however, that the value of $b$ has not been 
fitted to maximize the coincidence with the lattice results; $b$ has been 
simply introduced in order to demonstrate  a concrete (and minimal) realization of 
the proposed mechanism for getting a divergent $\lambda_{2}(p)$. 
Turning to the case of the  ``symmetric limit'', one observes (see Fig.~\ref{fig:lambda1-tau5_symm})
that our predictions follow rather accurately the pattern of the lattice data, but 
a coincidence may be achieved only after rescaling by a constant factor; specifically, 
$\lambda'_1(p)$ must be scaled down by a factor of 0.8, while 
$\tau_5(p)$ must be scaled up by a factor of 3.

Of course, it is clear
that we are far from having performed an exhaustive numerical study of
the theoretical quark-gluon vertex  solutions found.  Indeed, in order
to   do   that,   one    should   solve   the   system   composed   by
\2eqs{general-sol-L}{general-sol-T}  allowing  the  momenta $p_1$  and
$p_2$ to  be general,  and using an  iterative procedure of  which the
one-loop  dressed   approximation  used  here   represents  the  first
step.  After the  iterative solution  becomes stable,  one  would then
project to the  various momenta configurations (soft gluon, symmetric,
zero quark momentum) studied here and, at that point, possibly compare
to the lattice.  As this procedure is expected to distort the accidental angular
cancellations taking place for the one-loop dressed $\K{2}{}$, one expects
to find  directly the  $1/p^2$ divergence seen  on the lattice  in the
soft gluon limit. In  addition, as mentioned above, one should also be able to assess the
quality of the minimal  Ansatz of~\cite{Qin:2013mta}, which was readily
assumed for the transverse  form factors of the background quark-gluon
vertex $\widehat{\Gamma}$. We hope to address some of these points in the near future.

It would be certainly interesting to apply the results obtained here, 
and in particular the general solution presented in Sec.~\ref{sec:complete} C, 
to phenomenologically relevant situations. In particular, the 
quark-gluon vertex is an essential ingredient of the Bethe-Salpeter kernel 
that appears in the calculations of the hadronic spectrum by means of integral equations
~\cite{Maris:1999nt,Chang:2009zb,Williams:2014iea}.
Since, in this case, 
some of the momenta entering into the vertex are integrated over, 
one would have to develop the tools that allow the 
computation of the form factors \2eqs{general-sol-L}{general-sol-T} for arbitrary momentum configurations.
A preliminary step in this direction is already reported in Appendix~\ref{1ldints}; however, additional theoretical 
work is required, since, depending on the external kinematics,
 the integration momenta of the relevant Bethe-Salpeter equations 
are known to pass from the Euclidean to the Minkowski space, \linebreak see~\eg~\cite{Maris:2003vk,Meyers:2012ka}. 
It would be worthwhile to explore the possibilities that the present approach 
may offer for accomplishing this challenging endeavor.  

It  must be  clear from  the detailed  presentation and  the pertinent
comments  made  throughout  this  article that  the  proposed  method
incorporates ingredients gathered from a diverse variety of techniques
and  formalisms.  In  particular,  while  the  fundamental  symmetries
provide  the  starting  point  by  furnishing  a  minimal  Ansatz,  an
important  part  of  the   answer  originates  from  the  diagrammatic
calculation  of  the  special   three-point  function,  where  lattice
propagators  are  used  as  input.    In  that  sense,  the  practical
feasibility   of  the   method   and  its   potential  usefulness   in
phenomenological  applications   relies  heavily  on   the  judicious
combination of  all these ingredients into  a self-consistent picture.
This  particular task,  in turn,  requires a  coordinated effort  from
different sectors of the physics community (such as SDEs and lattice).
Despite  these  apparent  limitations,  in our  opinion  an  important
advantage of this method is that it provides a definite prediction for
{\it all}  twelve form  factors of the  quark-gluon vertex.  Given the
paramount  phenomenological importance  of  some of  them~\cite{Chang:2010hb}, the  effort invested  in  overcoming  the  aforementioned  difficulties might be particularly rewarding.

\acknowledgments 

The research of J.~P. is supported by the Spanish MEYC under 
grant FPA2011-23596. The work of  A.~C.~A  is supported by the 
National Council for Scientific and Technological Development - CNPq
under the grant 306537/2012-5 and project 473260/2012-3,
and by S\~ao Paulo Research Foundation - FAPESP through the project 2012/15643-1.

\appendix

\section{\label{app:NCandT+L}Relations between the NC and T+L bases}

The form factors of the $T+L$ basis are related to those of the naive conjugate basis through the relations~\cite{Davydychev:2000rt},
\begin{align}
%f1L
f^{\s{L}}_1&=f_1-\frac12(p_2\cd q)(f_6-f_7)-\frac12(p_1\cd q)(f_8-f_9)+(p_1\spr p_2)f_{12},\nonumber \\
%f2L
f^{\s{L}}_2&=\frac1{2(q\spr t)}\left[(p_2\cd q)(f_6+f_7)+(p_1\cd q)(f_8+f_9)\right],\nonumber \\
%f3L
f^{\s{L}}_3&=\frac1{q\spr t}\left\{(p_2\cd q)\left[f_2+(p_1\cd p_2)f_{10}\right]+(p_1\cd q)\left[f_3+(p_1\cd p_2)f_{11}\right]\right\},\nonumber\\
%f4L
f^{\s{L}}_4&=\frac12\left[f_4+f_5+(p_2\cd q)f_{10}+(p_1\cd q)f_{11}
\right],
\label{NaiveconjtoT+L:L} 
\end{align}
and
\begin{align}
%f1T
f^{\s{T}}_1&=\frac1{q\spr t}\left[f_2-f_3+(p_1\cd p_2)(f_{10}-f_{11})\right],\nonumber \\
%f2T
f^{\s{T}}_2&=\frac1{2(q\spr t)}\left[f_6+f_7-f_8-f_9\right],\nonumber \\
%f3T
f^{\s{T}}_3&=-\frac14\left[f_6-f_7-f_8+f_9
\right],\nonumber \\
%f4T
f^{\s{T}}_4&=\frac1{q\spr t}\left[f_{10}-f_{11}\right],\nonumber\\
%f5T
f^{\s{T}}_5&=-\frac12\left[f_4-f_5\right],\nonumber\\
%f6T
f^{\s{T}}_6&=\frac14\left[f_6-f_7+f_8-f_9\right],\nonumber 
\end{align}
\begin{align}
%f7T
f^{\s{T}}_7&=-\frac1{q\spr t}\left[(p_2\cd q)f_{10}+(p_1\cd q)f_{11}\right],\nonumber \\
%f8T
f^{\s{T}}_8&=f_{12}.
\label{NaiveconjtoT+L:T} 
\end{align} 
Conversely one has
\begin{align}
%f1
f_1&=f_1^{\s{L}}+q^2f_3^{\s{T}}+(q\spr t)f_6^{\s{T}}-(p_1\cd p_2)f_8^{\s{T}},\nonumber \\
%f2
f_2&=f_3^{\s{L}}+(p_1\cd q)f_1^{\s{T}}-(p_1\cd p_2)(p_1\cd q)f_4^{\s{T}}+(p_1\cd p_2)f_7^{\s{T}},\nonumber \\
%f3
f_3&=f_3^{\s{L}}-(p_2\cd q)f_1^{\s{T}}+(p_1\cd p_2)(p_2\cd q)f_4^{\s{T}}+(p_1\cd p_2)f_7^{\s{T}},\nonumber \\
%f4
f_4&=f_4^{\s{L}}-f_5^{\s{T}}+\frac12(q\spr t) f_7^{\s{T}},\nonumber \\
%f5
f_5&=f_4^{\s{L}}+f_5^{\s{T}}+\frac12(q\spr t) f_7^{\s{T}},\nonumber \\
%f6
f_6&=f_2^{\s{L}}+(p_1\cd q)f_2^{\s{T}}-f_3^{\s{T}}+f_6^{\s{T}},\nonumber \\
%f7
f_7&=f_2^{\s{L}}+(p_1\cd q)f_2^{\s{T}}+f_3^{\s{T}}-f_6^{\s{T}},\nonumber \\
%f8
f_8&=f_2^{\s{L}}-(p_2\cd q)f_2^{\s{T}}+f_3^{\s{T}}+f_6^{\s{T}},\nonumber \\
%f9
f_9&=f_2^{\s{L}}-(p_2\cd q)f_2^{\s{T}}-f_3^{\s{T}}-f_6^{\s{T}},\nonumber \\
%f10
f_{10}&=(p_1\cd q)f_4^{\s{T}}-f_7^{\s{T}},\nonumber \\
%f11
f_{11}&=-(p_2\cd q)f_4^{\s{T}}-f_7^{\s{T}},\nonumber \\
%f12
f_{12}&=f_8^{\s{T}}.
\label{T+LtoNaiveconj}
\end{align}

%%%%%%%%%%%%%%%%%%%%%%%%%%%%%%%%%%%%%%%%%%%%%%%%%%%%%%%%%%%%%%%%%%%%%%%%%%%%%%%%%%%%%%%%%%%%%%%%%%%%%%%%%%%%%%%

%%%%%%%%%%%%%%%%%%%%%%%%%%%%%%%%%%%%%%%%%%%%%%%%%%%%%%%%%%%%%%%%%%%%%%%%%%%%%%%%%%%%%%%%%%%%%%%%%%%%%%%%

\section{\label{1ldints}One-loop dressed integrals for general momenta}

In the case of arbitrary momenta $p_1$ and $p_2$, 
we split the one-loop dressed function $K^\mu$, defined in Eq.~(\ref{kkk}), according to
\be
K^\mu (q,p_2,-p_1) = \frac{i}{2}g^2C_A\underbrace{\int_k (\sla{k} + \sla{p}_2)\gamma^\nu P^\mu_\nu(k)\R{\s{A}}(k,p_1,p_2)}_{K_1^\mu(p_1,p_2)}+\frac{i}{2}g^2C_A\underbrace{\int_k \gamma^\nu P^\mu_\nu(k)\R{\s{B}}(k,p_1,p_2)}_{K_2^\mu(p_1,p_2)}, 
\ee
where we have defined 
\be \label{Rgeneral}
R^f(k,p_1,p_2) = \frac{f(k+p_2)\Delta(k^2)}{A^2(k+p_2)(k+p_2)^2-B^2(k+p_2)}D(k-q),
\ee
which reproduces \2eqs{Rf-soft-symm}{Reqzq} in the corresponding kinematic limits ($f=A,B$ as usual).

The objective is then to project out the above integrals such  
that they become expressed in terms of the tensors appearing in the naive conjugated basis. If we start with the integral $K_1^\mu$, since one has 
\be 
(\sla{k} + \sla{p}_2)\gamma^\nu P^\mu_\nu(k) = \sla{p}_2\gamma^\mu + \sla{k}\gamma^\mu - \frac{k^\mu}{k^2}[k\cdot(k+p_2)] - \frac{k^\mu}{k^2}\widetilde{\sigma}_{\rho\nu} p_2^\rho k^\nu,
\label{deckgP}
\ee
one may reorganize this integral in the form
\be \label{K1dec}
K_1^\mu(p_1,p_2) = \sum_{i=1}^4 \Jgen_i^\mu(p_1,p_2),
\ee
with
\begin{align}
\Jgen_1^\mu(p_1,p_2) &= \sla{p}_2\gamma^\mu \int_k \R{\s{A}}(k,p_1,p_2);& 
\Jgen_2^\mu(p_1,p_2) &= \int_k \sla{k}\gamma^\mu \R{\s{A}}(k,p_1,p_2); \nonumber \\
\Jgen_3^\mu(p_1,p_2) &= -\int_k \frac{k^\mu}{k^2}[k\cdot(k+p_2)] \R{\s{A}}(k,p_1,p_2);& 
\Jgen_4^\mu(p_1,p_2) &= -\widetilde{\sigma}_{\rho\nu} p_2^\rho\int_k \frac{k^\mu k^\nu}{k^2} \R{\s{A}}(k,p_1,p_2).
\label{JK1int}
\end{align}
It is then straightforward to show that
\begin{align} 
\Jgen_1^\mu &= \Igen_1^\s{A}(\N{2}^\mu - \N{4}^\mu),\nonumber  \\
\Jgen_2^\mu &= \frac{1}{r}\left\{ \left[p_1^2\Igen_2^\s{A} - (p_1\spr p_2)\Igen_3^\s{A}](\N{2}^\mu - \N{4}^\mu) + [p_2^2\Igen_3^\s{A} - (p_1\spr p_2)\Igen_2^\s{A}\right](\N{3}^\mu - \N{5}^\mu)\right\}, \nonumber \\ 
\Jgen_3^\mu &= -\frac{1}{r} \left\{\left[p_1^2(\Igen_2^\s{A} + \Igen_5^\s{A}) - (p_1\spr p_2)(\Igen_3^\s{A} + \Igen_4^\s{A})\right]\N{2}^\mu\right. \nonumber \\
&\left.+ \left[p_2^2(\Igen_3^\s{A} + \Igen_4^\s{A}) - (p_1\cdot p_2)(\Igen_2^\s{A} + \Igen_5^\s{A})\right]\N{3}^\mu \right\}, 
\nonumber \\
\Jgen_4^\mu &= \frac{M^\s{A}}{2r}\N{4}^\mu + \frac{1}{r}\left[p_2^2\Igen_1^\s{A}  - \Igen_5^\s{A}  - \frac{3p_2^2}{2r}M^\s{A} \right]\left[\N{11}^\mu-(p_1\spr p_2)\N{3}^\mu\right]\nonumber \\
&+ \frac{1}{r}\left[(p_1\spr p_2)\Igen_1^\s{A}  - \Igen_4^\s{A}  - \frac{3(p_1\spr p_2)}{2r}M^\s{A} \right]\left[(p_1\spr p_2)\N{2}^\mu-\N{10}^\mu\right],
\end{align}
where we have set
\begin{align}
\Igen_1^f(p_1,p_2) &= \int_k \R{f}(k,p_1,p_2); &
\Igen_2^f(p_1,p_2) &= \int_k (k\spr p_2)\R{f}(k,p_1,p_2), \nonumber \\ 
\Igen_3^f(p_1,p_2) &= \int_k (k\spr p_1)\R{f}(k,p_1,p_2); &
\Igen_4^f(p_1,p_2) &= \int_k \frac{(k\spr p_1)(k\spr p_2)}{k^2}\R{f}(k,p_1,p_2), \nonumber \\ 
\Igen_5^f(p_1,p_2) &= \int_k \frac{(k\spr p_2)^2}{k^2}\R{f}k,p_1,p_2);&
\Igen_6^f(p_1,p_2) &= \int_k \frac{(k\spr p_1)^2}{k^2}\R{f}(k,p_1,p_2), \nonumber \\ 
M^f(p_1,p_2) &= r\Igen_1^f + 2(p_1\spr p_2)\Igen_4^f - p_1^2\Igen_5^f - p_2^2\Igen_6^f.
\end{align}

For the integral $K_2^\mu$ we may instead write
\be 
K_2^\mu(p_1,p_2) = \Jgen_5^\mu(p_1,p_2) + \Jgen_6^\mu(p_1,p_2),
\ee
with
\begin{align}
\Jgen_5^\mu(p_1,p_2) &= \gamma^\mu \int_k \R{\s{B}}(k,p_1,p_2); &
\Jgen_6^\mu(p_1,p_2) &= -\gamma_\nu \int_k \frac{k^\mu k^\nu}{k^2} \R{\s{B}}(k,p_1,p_2).
\end{align}
It is then immediate to show that
\begin{align}
\Jgen_5^\mu &= \Igen_1^\s{B} \N{1}^\mu, \nonumber \\
\Jgen_6^\mu &= -\frac{M^\s{B}}{2r}\N{1}^\mu + \frac{1}{r}\left[(p_1\spr p_2)\Igen_1^\s{B} - \Igen_4^\s{B} - \frac{3(p_1\spr p_2)}{2r}M^\s{B}\right] (\N{7}^\mu + \N{8}^\mu) \nonumber \\
&+ \frac{1}{r}\left[\frac{3p_1^2}{2r}M^\s{B} + \Igen_6^\s{B} - p_1^2\Igen_1^\s{B}\right] \N{6}^\mu + \frac{1}{r}\left[\frac{3p_2^2}{2r}M^\s{B} + \Igen_5^\s{B} - p_2^2\Igen_1^\s{B}\right] \N{9}^\mu.
\end{align}

Using the results above one can therefore recover all the twelve form factors characterizing $K^\mu$ in the NC basis; the corresponding expressions in the T+L basis can be obtained by using \2eqs{NaiveconjtoT+L:L}{NaiveconjtoT+L:T}.

%\bibliography{../../../../../Bibliography/bibliography}

\end{document}